\documentclass{article}
\usepackage[T1]{fontenc}
\usepackage{lmodern}
\usepackage{csquotes}
\usepackage{amsmath,amssymb,amsthm}
\usepackage{mathtools}
\usepackage[table]{xcolor}
\usepackage{array,booktabs}
\usepackage{tikz}
\usepackage{braket}
\usepackage{microtype}
\usepackage{xurl}
\usepackage{hyperref}
\usepackage{cleveref}

\newtheorem{lemma}{Lemma}
\newtheorem{theorem}{Theorem}

\newtheorem{corollary}{Corollary}
\newtheorem*{remark}{Remark}
\usepackage{placeins}

\usetikzlibrary{shapes,arrows.meta,positioning,calc,intersections,shapes.geometric}

\usepackage{biblatex}
\graphicspath{ {./figures/} }
\newcommand{\bigO}[1]{\mathcal{O}\!\left(#1\right)}

\newcommand{\void}[1]{}
\newcommand{\sel}{\mathsf{S}}

\newcommand{\branch}{\mathsf{B}}

\title{Trading Memory for Depth: Tight Bounds for LCU Block-Encodings}
\author{Sebastian Issel\\
  \small Fraunhofer Institute for Applied and Integrated Security AISEC\\
  \small Lichtenbergstra{\ss}e 11, 85748 Garching, Germany}
\date{November 2026}

\begin{document}
\maketitle

\begin{abstract}
An LCU block-encoding of \(A=\sum_j\alpha_j U_j\) runs the \(U_j\) in
series on one register, so its select depth is the total work
\(\sum_j d_j\). Depth was the one block-encoding constraint with no knob:
grouping the \(T\) terms into \(g\) registers that run in parallel gives it
one, and the binning is the knob. For a \(g\)-branched realisation (up to
\(g\) oracles in parallel) the select depth is tight at
\(\Theta(C+\log(g\,\max\{n,m\}))\) for \(g\ge2\), with \(C\) the binning
makespan and \(m\) the largest bin, in the \(k\)-local atomic-oracle model.
The work floor is a two-way reduction to classical multiprocessor
scheduling: every realisation \emph{induces} a \(g\)-machine schedule, and the
construction realises any binning, so an optimal binning attains it. The
additive logarithmic routing term is a light-cone cost, independent of the
selector encoding and robust to register reuse, measurement, reset, and
teleportation; width is at least \(gn\). The standard \(1/\alpha\)
subnormalisation and success probability are unchanged; the endpoints are
sequential LCU (\(g{=}1\)) and the fully parallel realisation (\(g{=}T\)).
Allowing the oracles to be split (McNaughton's rule) reaches the work floor
\(\sum_j d_j/g\). Two selector realisations trade space for depth: a dense
(M\"ott\"onen) one is memory-tight at the standard LCU footprint but its
multiplexed controls cost a \(\log\log(m{+}1)\) factor on the makespan; a
one-hot one is depth-tight at \(T\) selector qubits. Neither reaches both
at once.
\end{abstract}

\section{Introduction}
\label{sec:intro}

The linear combination of unitaries (LCU) block-encodes a general operator
\(A=\sum_j \alpha_j U_j\) on a quantum computer, in the standard
setting of \textcite{Gilyen_2019}; we additionally track the \(k\)-local depth
\(d_j\) of each term, which the standard treatment leaves unspecified.
The recipe is to prepare a \emph{selector}
state
\(\ket{\phi}=\sum_j\sqrt{\lvert\alpha_j\rvert/\alpha}\,\ket j\), apply the
\(U_j\) controlled on it, and postselect, yielding a \(1/\alpha\)-block-encoding
of \(A\).

Its select step is a constraint, like the others. A block-encoding is
usable only when its constraints all hold: the \(1/\alpha\)
subnormalisation (hence the success probability and the amplification cost),
the width, the error budget, and the depth. The first three could be traded;
the fourth could not. A standard select multiplexes the \(U_j\) onto a single
data register, so they run in series and the depth is the total work
\(\sum_j d_j\), with no dependence on the depth \(\max_j d_j\) of the deepest
single term. The select had no parameter to tune, so depth sat outside the
tradeoff space the other constraints occupied. We settle how much of it extra
memory buys back, and where the wall sits.

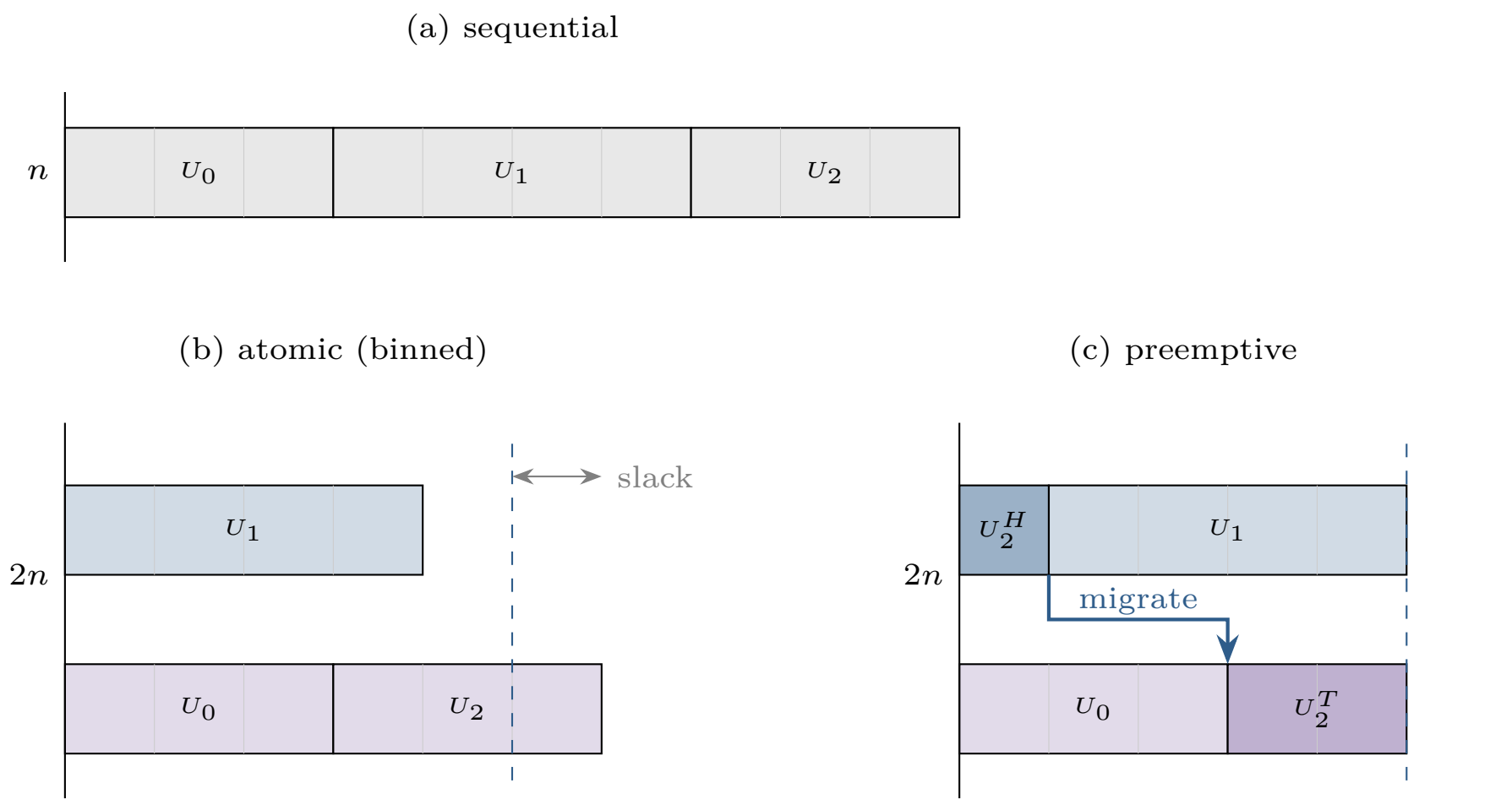
\begin{figure}[t]
\centering
\begin{tikzpicture}[>=Stealth, x=7mm, y=7mm,
  ujfill/.style ={draw, fill=gray!18},
  ujBfill/.style={draw, fill=gray!40},
  regA/.style ={draw, fill=accent!22},
  regAmig/.style={draw, fill=accent!48},
  regB/.style ={draw, fill=select!22},
  regBmig/.style={draw, fill=select!48},
  reg/.style={font=\scriptsize},
  ax/.style={font=\scriptsize},
  floorline/.style={dashed, accent},
  boxh/.style={font=\tiny},
]
\definecolor{accent}{HTML}{2E5C8A}
\definecolor{select}{HTML}{7A5C9E}
\newcommand{\DepthBox}[5]{%
  \draw[#1] (#2,#4-0.5) rectangle (#2+#3,#4+0.5);
  \pgfmathtruncatemacro{\nunits}{#3-1}%
  \ifnum\nunits>0
    \foreach \i in {1,...,\nunits} {
      \draw[gray!40, very thin] (#2+\i,#4-0.5) -- (#2+\i,#4+0.5);
    }
  \fi
  \node[boxh] at (#2+#3/2,#4) {#5};%
}

\begin{scope}[shift={(0,0)}]
  \DepthBox{ujfill}{0}{3}{1}{$U_0$}
  \DepthBox{ujfill}{3}{4}{1}{$U_1$}
  \DepthBox{ujfill}{7}{3}{1}{$U_2$}
  \draw (0,0) -- (0,1.9); \node[reg,left] at (0,1) {$n$};
  \node[ax] at (5,2.6) {(a) sequential};
\end{scope}

\begin{scope}[shift={(0,-6)}]
  \DepthBox{regA}{0}{4}{3}{$U_1$}
  \DepthBox{regB}{0}{3}{1}{$U_0$}
  \DepthBox{regB}{3}{3}{1}{$U_2$}
  \draw[floorline] (5,0.2)--(5,4.0);
  \draw (0,0) -- (0,4.2); \node[reg,left] at (0,2.5) {$2n$};
  \node[ax] at (3,5.0) {(b) atomic (binned)};
  \draw[<->,gray] (5,3.6)--(6,3.6) node[right,ax,gray] {slack};
\end{scope}

\begin{scope}[shift={(10,-6)}]
  \DepthBox{regAmig}{0}{1}{3}{$U_2^H$}
  \DepthBox{regA}{1}{4}{3}{$U_1$}
  \DepthBox{regB}{0}{3}{1}{$U_0$}
  \DepthBox{regBmig}{3}{2}{1}{$U_2^T$}
  \draw[floorline] (5,0.2)--(5,4.0);
  \draw (0,0) -- (0,4.2); \node[reg,left] at (0,2.5) {$2n$};
  \node[ax] at (2.5,5.0) {(c) preemptive};
  \draw[->, thick, accent] (1,2.5) -- (1,2.0)
    -- (3,2.0) node[ax, midway, above, yshift=-2.5pt] {migrate}
    -- (3,1.5);
\end{scope}

\end{tikzpicture}
\caption{Space--time diagrams (depth vs.\ qubit width) for three terms
$d=(3,4,3)$, $g{=}2$: (a)~sequential, depth $\sum_j d_j=10$; (b)~atomic
space--time LCU, makespan $C^*=6$ (bin-packing slack); (c)~preemptive, work
floor $\sum_j d_j/g=5$. (c)~reaches the floor, (b) exceeds it by its slack
(\cref{sec:schedule}).}
\label{fig:spacetime}
\end{figure}

Group the \(T\) terms into \(g\) bins. A family (\emph{space--time LCU}),
one realisation per binning and indexed by \(g\in[1,T]\), realises \emph{any}
binning at depth
\[ D\;=\;C+\bigO{\log(g\,\max\{n,m\})}, \]
where \(C\) is the realised binning makespan and \(m\) the number of terms in
the largest bin; \(g{=}1\) and \(g{=}T\) recover sequential and fully parallel
LCU. The bound is tight: in the \(k\)-local atomic-oracle model the
construction is matched by a lower bound
\(C^{*}+\Omega(\log(g\,\max\{n,m\}))\)
(the work floor plus a logarithmic routing term the light cone forces),
up to the bin-packing approximation. The work floor is a two-way reduction to
classical multiprocessor scheduling: every \(g\)-branched realisation
\emph{induces} a \(g\)-machine schedule, so the quantum depth is bounded below
by the classical makespan
\(C^{*}\ge\max\{\max_j d_j,(\sum_j d_j)/g\}\), and the construction realises
any binning, so an optimal binning attains it. Register reuse, measurement,
reset, and teleportation do not help, since they only relocate the same work.
Allowing the oracles to be non-atomic turns the binning into preemptive
scheduling and reaches the work floor \(\sum_j d_j/g\) (McNaughton's rule,
\cref{sec:schedule}). The operating point is a binning (\cref{fig:spacetime}).

The construction realises the bound by distributing, not copying, the input
over the branch registers; the oracles run in parallel on their parts and the
inverse routing recombines the branches before the postselection forms the sum.
Only this postselection is non-unitary. The postselected data register carries
precisely the state a sequential LCU of the same terms would produce, so the
encoding is the standard LCU one: same \(1/\alpha\) subnormalisation, success
probability, and postselected state, and amplitude amplification composes
unchanged.

Two selector realisations trade space for depth (\cref{sec:iterate},
\cref{tab:tradeoff}). The dense (M\"ott\"onen) realisation is memory-tight at
the standard LCU footprint \(\sum_k\lceil\log_2(m_k{+}1)\rceil\), but its
multiplexed oracle controls cost a \(\log\log(m{+}1)\) factor on the makespan,
depth \(C\log\log(m{+}1)+\bigO{\log(g\,\max\{n,m\})}\). The one-hot realisation
is depth-tight, \(C+\bigO{\log(g\,\max\{n,m\})}\) (single-qubit controls), at
the cost of \(T\) selector qubits. Neither reaches both at once.

\paragraph{Contributions.}
\begin{itemize}
  \item \textbf{Tight bounds.} The select depth is
        \(\Theta(C+\log(g\,\max\{n,m\}))\) for \(g\ge2\)
        (\(g{=}1\) sequential), matched by a lower bound
        \(C^{*}+\Omega(\log(g\,\max\{n,m\}))\) in the \(k\)-local atomic-oracle
        model, robust to register reuse, measurement, reset, and teleportation
        (\cref{thm:hybrid,thm:pointwise}).
  \item \textbf{The reduction.} The work floor is a two-way reduction to
        classical multiprocessor scheduling: every \(g\)-branched realisation
        \emph{induces} a \(g\)-machine schedule (\cref{lem:induced}), and the
        construction realises any binning, so an optimal binning attains the
        bound; the routing term is a \(k\)-local light-cone cost
        (\cref{lem:routing,lem:withbin}).
  \item \textbf{The construction} (\cref{sec:iterate}): a family of
        block-encodings, one per binning, at the tight depth above, with the
        standard \(1/\alpha\) subnormalisation and success probability. Two
        selector realisations trade space for depth: a dense (M\"ott\"onen) one,
        memory-tight but with a \(\log\log(m{+}1)\) depth overhead
        (\cref{prop:compress}); and a one-hot one, depth-tight at \(T\) selector
        qubits (\cref{thm:hybrid}).
  \item \textbf{Preemptive variant} (\cref{sec:schedule}): when the oracles can
        be interrupted and resumed, cutting terms and migrating the pieces
        reaches the work floor \(\sum_j d_j/g\) exactly, at \(g-1\) cuts.
\end{itemize}

\section{Setup and notation}
\label{sec:setup}

Let \(U_0,\dots,U_{T-1}\) be unitaries acting on \(n\) qubits
and let \(\alpha_j\in\mathbb C\) be complex coefficients, written
\(\alpha_j=\lvert\alpha_j\rvert e^{i\theta_j}\).
We write
\[
  A \;=\; \sum_{j=0}^{T-1} \alpha_j U_j,
  \qquad
  \alpha \;=\; \sum_{j=0}^{T-1} \lvert\alpha_j\rvert
\]
for the operator to be block-encoded and its subnormalisation \(\alpha\), the
\(\ell_1\)-norm of the coefficients, and
\[
  \ket{\phi} \;=\; \sum_{j=0}^{T-1} \sqrt{\lvert\alpha_j\rvert/\alpha}\,\ket{j}
\]
for the normalised \emph{selector state}. When the terms are grouped into
\(g\) bins of sizes \(m_0,\dots,m_{g-1}\) (\(\sum_k m_k=T\), fixed in
\cref{sec:iterate}), we write \(m=\max_k m_k\) for the largest bin. Its concrete realisation
(a binary address on \(\lceil\log_2 T\rceil\) qubits in standard LCU, or a
one-hot encoding on \(T\) qubits in our alternative construction, \cref{sec:iterate})
is a design choice fixed below. The phase \(e^{i\theta_j}\) is not
in the amplitude (which is real) but absorbed into the controlled oracle,
applied as the phased \(e^{i\theta_j}U_j\) in sector \(j\)
(\cref{sec:gadget}). We assume
the oracles are provided as \emph{controlled}\,\(U_j\) (the standard LCU
convention), and write \(d_j\) for the \(k\)-local depth of
controlled-\(U_j\); higher control (steering \(U_j\) from an \(r\)-qubit
address) costs \(\bigO{\log r}\) depth under bounded fan-out with
\(\bigO{r}\) borrowed ancilla (the decomposition used for the dense
realisation, \cref{rem:ancilla}); ancilla-free it costs \(\bigO{r}\).
The goal is a circuit that, on a \(n\)-qubit input \(\ket{\psi}\),
produces a state proportional to \(A\ket{\psi}\), i.e.\ an
\(\bigO{1/\alpha}\)-block-encoding of \(A\).
All logarithms are base two.

\paragraph{Gate model.}\label{sec:model}
We work in the \emph{bounded-arity} (\(k\)-local) circuit model: each gate acts
on at most \(k=\bigO{1}\) qubits, and \emph{depth} is the minimum number of
layers of pairwise-disjoint gates (a qubit participates in at most one gate
per layer, even as a read-only control). This is the standard, physical model
for block-encoding cost accounting. The alternative of unbounded-arity gates is
a non-physical idealisation under which an \(\bigO{Tn}\)-qubit unitary is one
gate and the routing terms disappear, leaving only the oracle work
\(\max_j d_j\); we state each depth lower bound in the \(k\)-local model.
Mid-circuit measurement, reset, and teleportation do not improve the bounds:
a running oracle's \(n\)-qubit intermediate state cannot be shrunk by relocating
it (\cref{sec:bound}).

\section{A depth lower bound}
\label{sec:bound}

Fix a binning of the \(T\) terms into \(g\) branches (the construction that
realises any binning is in \cref{sec:iterate}); we argue over it classically.
The reduction runs both ways: every width-\(gn\) realisation \emph{induces}
a \(g\)-machine schedule (\cref{lem:induced}), so the quantum depth is bounded
below by the classical makespan, and the construction realises any binning, so
an optimal one attains it. The matching routing term is structural in the
binning's address space (\(g\) branches, \(m\) terms in the largest bin, \(n\)
data bits), independent of how the selector is encoded.

\paragraph{The computation window.}
To apply \(U_j\), an \(n\)-qubit controlled oracle of \(k\)-local depth
\(d_j\), the circuit dedicates \(n\) qubits to \(U_j\)'s evolving intermediate
state for \(d_j\) consecutive layers: they cannot be reset mid-computation
without destroying it. This \(n\)-wide window is the oracle's interface width,
not a density claim about \(U_j\)'s internals; an opaque atomic oracle cannot
be parallelised by inspecting its interior, so two windows cannot share a
register concurrently under \(k\)-locality. The \(k\)-local disjoint-gate
convention (a qubit is in at most one gate per layer, even as a control) makes
concurrent calls qubit-disjoint, so at width \(gn\) at most \(g\) windows
overlap any layer. The window is robust: measurement, reset, and in-circuit
teleportation extend to a unitary on a larger (purified) system to which the
same argument applies, so the concurrency is not reduced (\cref{app:lem-induced}).
The opaque atomic oracle is the standard LCU setting (the \(U_j\) supplied as
controlled black boxes); relaxing to decomposable (circuit) oracles permits
preemption (\cref{sec:schedule}) but dissolves the non-preemptive guarantees.

\begin{lemma}[Width floor]
\label{lem:width}
Any realisation allowing \(g\) concurrent \(n\)-wide oracle calls has width
at least \(gn+\textstyle\sum_k\lceil\log_2(m_k{+}1)\rceil\). The data floor
\(gn\) is the \(g\) overlapping \(n\)-qubit oracle windows (a running oracle's
intermediate state occupies \(\ge n\) qubits for \(d_j\) layers wherever it is
parked, so \(g\) windows need \(\ge gn\); register reuse, measurement, reset,
and teleportation do not reduce it). The selector floor
\(\sum_k\lceil\log_2(m_k{+}1)\rceil\) is forced by the disjoint-gate model:
the \(g\) branch oracles run concurrently, each controlled on the selector, so
a selector qubit shared across branches would control \(g\) gates in one layer,
violating disjoint-gate parallelism; hence the selector splits into \(g\)
disjoint per-branch parts, and branch \(k\) must encode its \(m_k\) terms plus
the reserved ``off'' state (at most one branch is active), needing
\(\ge\lceil\log_2(m_k{+}1)\rceil\) qubits.
\end{lemma}

This single fact, that each \(U_j\) is an \(n\)-wide window of duration \(d_j\)
with at most \(g\) overlapping at width \(gn\), is the work floor
(\cref{lem:induced}); the routing lower bound adds a light-cone term to it
(\cref{lem:routing,lem:withbin}).

\begin{lemma}[Induced schedule]
\label{lem:induced}
In the \(k\)-local disjoint-gate model, any atomic (non-preemptive) LCU
realisation at width \(gn\) --- opaque black-box oracles, the standard LCU
setting --- induces a \(g\)-machine non-preemptive schedule of the \(T\) jobs
\(\{d_j\}\); hence its depth satisfies \(D\ge C^*\), the optimal such makespan.
Under the stronger decomposable-oracle model (each \(U_j\) a circuit, invokable
as pieces of total depth \(d_j\); \cref{sec:schedule}), any preemptive
realisation induces a \(g\)-machine preemptive schedule, hence
\(D\ge\max\{\max_j d_j,\,\tfrac{1}{g}\textstyle\sum_j d_j\}\), the Brent floor.
\end{lemma}
\begin{proof}[Proof sketch]
Each \(U_j\)'s gates occupy definite circuit layers (the bound is on layout
width, not branch amplitudes). Let \(I_j\subset[0,D)\) be the layer interval of
the call to \(U_j\); at most \(g\) intervals overlap any layer, since each
window is \(n\)-wide against width \(gn\). For atomic calls the family has
pointwise depth \(\le g\), hence is \(g\)-colourable (interval graphs colour
greedily); a colour is a machine, so the induced schedule is a \(g\)-machine
non-preemptive one of makespan \(\le D\), giving \(D\ge C^*\). The preemptive
case gives the Brent floor \(\max\{\max_j d_j,\sum_j d_j/g\}\) from the same
concurrency bound and causal ordering of one job's pieces.
\Cref{app:proofs} gives the full argument, including the \(k\)-local robustness
clause.
\end{proof}

\paragraph{Routing: a prep lower bound.}
Routing is a lower bound on the \emph{preparation} that precedes the oracle,
not a separate cost added to it. In the \(k\)-local model a circuit spanning
\(t\) target qubits from one source has light cone \(k^d\), so needs
\(d\ge\log_k t=\Omega(\log t)\) depth. The prep must simultaneously span two
spaces, on disjoint registers, so the two costs combine by a max:

\begin{lemma}[Selector-prep span]
\label{lem:withbin}
Before any term oracle runs, the prep must build a selector that distinguishes
the \(g\cdot m\) term-addresses (\(g\) branches, \(m\) terms in the largest):
a state carrying \(\log(gm)\) bits of amplitude on the selector register. Under
bounded fan-out preparing such a state costs \(\Omega(\log(gm))
=\Omega(\log g+\log m)\). It is additive with the makespan: a
makespan-determining branch \(k^*\)'s oracle \emph{reads} its selector as a
control, so the prep that writes it must precede that oracle and conflicts with
it on the selector qubits. Independent of the selector encoding (one-hot, dense,
or joint): this is the multiplexed-select cost every LCU pays, sequential
included \parencite{Mottonen_2005,Gui_2023}.
\end{lemma}

\begin{lemma}[Data-prep span]
\label{lem:routing}
Before the oracles run, the \(n\)-qubit input must be made available on the
\(g\) branch registers: a map whose support is \(gn\) data qubits wide.
Routing one \(n\)-qubit state to \(g\) destinations under bounded fan-out
fans one selection signal to \(g\) targets (\(\Omega(\log g)\)) and moves
\(n\) data qubits under that signal (\(\Omega(\log n)\), a \(1\to n\)
fan-out), the move consuming the fanned signal so the two serialize, giving
\(\Omega(\log g+\log n)=\Omega(\log(gn))\). Additive with the makespan by the
same register-conflict argument: the distribution touches the makespan branch
\(k^*\)'s data register, which its oracle then occupies, so the two are on
disjoint layer intervals. Holds for any selector encoding; reset, measurement,
and teleportation extend to a unitary on a purified system and do not remove
the register conflict.
\end{lemma}

The two spans act on disjoint registers (selector vs. data), so they run in
parallel and only the longer survives, giving the routing depth
\(\max\{\Omega(\log(gm)),\,\Omega(\log(gn))\}=\Omega(\log(g\,\max\{n,m\}))\)
(\(\max\{\log a,\log b\}=\log\max\{a,b\}\); \cref{app:thm-pointwise}).
Both spans are robust in the \(k\)-local atomic-oracle model (measurement,
reset, and teleportation extend to a unitary on a larger system, so the
light-cone bound survives). Under the non-physical unbounded-arity idealisation
fan-out is a single gate and the routing terms disappear, leaving only the
oracle work --- the physical \(k\)-local model is where they are incurred.

\begin{theorem}[Pointwise depth lower bound]\label{thm:pointwise}
Fix a binning into \(g\ge2\) bins of max size \(m\). In the \(k\)-local
atomic-oracle model at width \(gn\), any realisation has
\[D\ge C^*+\max\!\Bigl\{\Omega(\log(gm)),\;\Omega(\log(gn))\Bigr\}
   =C^*+\Omega\!\bigl(\log(g\,\max\{n,m\})\bigr),\]
the work floor (\cref{lem:induced}) plus the prep spans of
\cref{lem:withbin,lem:routing}, additive with the makespan and combined by a
max over disjoint registers. The construction (\cref{thm:hybrid}) attains
\(C+\bigO{\log(g\,\max\{n,m\})}\), so the bound is \(\Theta\)-tight (exactly on
the routing, up to the bin-packing slack \(C-C^*\)). At \(g{=}1\) the
selector span \(\Omega(\log T)\) recovers sequential LCU; \(g{=}T, m{=}1\) is
the fully parallel realisation (\cref{thm:routing}).
\end{theorem}
\begin{proof}[Proof sketch]
\Cref{lem:induced} gives \(D\ge C^*\). The prep precedes the makespan branch's
oracle (which reads the registers the prep writes), so it is additive:
\cref{lem:routing} gives \(D\ge C^*+\Omega(\log(gn))\) and \cref{lem:withbin}
gives \(D\ge C^*+\Omega(\log(gm))\). The two spans act on disjoint registers
(data vs.\ selector), so they combine by a max, yielding
\(\log(g\,\max\{n,m\})\). The construction (\cref{thm:hybrid}) attains
\(C+\bigO{\log(g\,\max\{n,m\})}\), so upper and lower bounds match up to
constants and the bin-packing slack. \Cref{app:proofs} details the
\(\Theta\)-equivalence of the max.
\end{proof}

\begin{corollary}[Preemptive relaxation]\label{cor:preemptive-lb}
Under the stronger oracle model in which each \(U_j\) is decomposable into
slices (a circuit oracle, not an opaque black box; \cref{sec:schedule}), the
work floor relaxes from the non-preemptive makespan \(C^*\) to the Brent
floor \(\max\{\max_j d_j,\sum_j d_j/g\}\). The routing term is unchanged:
the prep spans of \cref{lem:withbin,lem:routing} are bounds on the preparation
that must precede the oracle, and slicing the oracle does not reduce the
\(gm\)-address selector space or the \(gn\)-qubit data space the prep must
span. Hence the preemptive lower bound is the Brent floor plus the same
\(\Omega(\log(g\,\max\{n,m\}))\).
\end{corollary}

A pre-shared entangled resource, as in the measurement-based parallelisation of
\textcite{Broadbent_2009}, can lower the routing depth, but only
by offloading it to a resource whose own preparation obeys the same light-cone
bound, so the trade is routing-depth for entanglement-preparation depth, not a
free reduction. The flat construction attains the flat bound:
\(D_{\parallel}=\max_j d_j+\bigO{\log T}\) (non-physical unbounded fan-out) or
\(+\bigO{\log(Tn)}\) (bounded, the physical model, via the selector spread of \cref{sec:iterate}), so
the routing overhead is \(\Theta\)-tight; parallelism costs exactly a
logarithmic routing term on top of the unavoidable work \(\max_j d_j\).

\begin{theorem}[Atomic tightness: the bridge]
\label{thm:bridge}
For any binning \(B\) the atomic construction (\cref{thm:hybrid}) tracks it:
\[
  D(B)=C(B)+\Theta\!\left(\log(g\,\max\{n,m\})\right)
  \quad(g\ge2),
\]
meeting the lower bound \cref{thm:pointwise},
\(D\ge C^*+\Omega(\log(g\,\max\{n,m\}))\), \(\Theta\)-tightly
when \(B\) is optimal (\(C(B)=C^*\); up to the bin-packing approximation
that LPT realises); at \(g{=}1\) the routing reduces to the
within-bin select (\cref{thm:pointwise}), \(D=C+\Theta(\log T)\). A suboptimal \(B\) leaves the
bin-packing slack \(C(B)-C^*\), a purely classical quantity. The gap from
\(C^*\) to the Brent floor \(\sum_j d_j/g\) is the bin-packing slack as a
phenomenon, closed by the preemptive variant (\cref{thm:preemptive}).
\end{theorem}
\begin{proof}[Proof sketch]
The upper bound is \cref{thm:hybrid}:
\(D\le C(B)+\bigO{\log(g\,\max\{n,m\})}\); the lower bound is
\cref{thm:pointwise}:
\(D\ge C^*+\Omega(\log(g\,\max\{n,m\}))\). They match up to constants when
\(C(B)=C^*\), i.e.\ under an optimal (or LPT-approximate) binning; the
difference \(C(B)-C^*\) is pure bin-packing slack. The lower bound is the
reduction of \cref{lem:induced} (work floor) plus the routing of
\cref{lem:routing,lem:withbin}. \Cref{app:proofs} gives the full derivation.
\end{proof}

\begin{remark}[Minimal area]
\label{cor:upper}
Area (qubit-time) is width times depth. The width floor is
\(\Omega(gn+\textstyle\sum_k\lceil\log_2(m_k{+}1)\rceil)\)
--- the \(gn\) branch registers forced by the \(g\)
concurrent \(n\)-wide oracle windows, and the selector term carrying the
\(m\) dependence (\(\sum_k\log(m_k{+}1)\le g\log m\)) --- and
the depth floor is
\(C+\Theta\!\left(\log(g\,\max\{n,m\})\right)\) (\cref{thm:bridge}), which also
contains \(\log m\). Their product is thus an area lower bound. It is
quasi-attained: the one-hot realisation reaches the depth floor (\cref{thm:hybrid})
but at width \(gn{+}T\), and the dense realisation (\cref{prop:compress}) reaches
the width floor but at depth \(C\,\log\log(m{+}1)\), a \(\log\log(m{+}1)\) factor above.
Both axes are thus reached up to the single \(\log\log(m{+}1)\) factor that the
dense multiplexed control already pays in standard LCU. Whether a single
construction can attain both axes simultaneously --- closing the area gap
entirely --- is open.
\end{remark}

\begin{remark}[Routing depth at the flat endpoint]
\label{thm:routing}
At the flat endpoint \(g{=}T, m{=}1\), \cref{thm:pointwise} specialises
to the bound on a \(g\)-branch parallel LCU:
\(\max_j d_j+\Omega(\log g+\log n)=\max_j d_j+\Omega(\log(gn))\) under bounded
fan-out (the within-bin select \(\Omega(\log m)\) vanishes with
\(m{=}1\), so the max collapses to the data-routing chain
\(\Omega(\log g+\log n)\)), which the construction attains.
\end{remark}

\section{The two-layer construction}
\label{sec:iterate}
\label{sec:gadget}

This section gives the construction that realises any binning at the tight
depth (the upper-bound half of the reduction in \cref{sec:bound}).

\paragraph{The controlled move.}
The construction rests on a single primitive, the \emph{controlled move}.
On a control qubit \(c\), a source register \(r\) holding \(\ket{x}\), and a
destination register \(s\) starting in \(\ket{0}\),
\[
  \mathrm{Move}(c;r\to s)
  \;=\;
  \mathrm{Toffoli}(c,s;r)\,\mathrm{Toffoli}(c,r;s),
\]
i.e.\ a controlled copy \(r\to s\) followed by a controlled erase
\(s\to r\), both controlled on \(c\). On basis states this sends
\(\ket{c}\ket{x}_r\ket{0}_s\mapsto\ket{c}\ket{0}_r\ket{x}_s\) when \(c=1\)
and fixes the registers when \(c=0\): a clean move that returns the source to
\(\ket{0}\), letting a single register serve as both input and output without
a separate data register. It is unitary, and because the one-hot selector
(defined below) keeps at most one branch live per sector, moves on disjoint
branch registers never collide.

\paragraph{The two-layer butterfly.}
We build the construction for \(T\) terms directly. It is a \emph{two-layer
one-hot butterfly}: an outer butterfly prepares a one-hot over the \(g\) blocks
(branches), and an inner butterfly per block splits each block's flag in place
into the \(m_k\) one-hot term flags. The two layers are separated by a routing
phase that moves the input to the active branch. The construction is
\(\mathrm{unprep}\circ\mathrm{oracle}\circ\mathrm{prep}\), where
\(\mathrm{prep}\) prepares the selector and routes the input,
\(\mathrm{oracle}\) applies the \(U_{k,l}\) as singly-controlled gates in
parallel, and \(\mathrm{unprep}=\mathrm{prep}^{\dagger}\). The only
non-unitary operation is the final postselection that forms the sum.
\emph{The routing is a max, not a sum.} The within-bin select (inner butterfly,
on the selector flags) and the data-routing chain (tree-move, on the branch
registers) act on disjoint qubit sets and share only the block-active signal, so
they run in parallel and only the longer survives. A one-qubit-per-block copy
\(\sel\to\sel^{\mathrm{copy}}\) of the block flags decouples them: it holds
the block flags intact through the inner butterfly (which destroys \(s_k[0]\)),
letting the path-decision flags be derived from \(\sel^{\mathrm{copy}}\) while
the inner runs on \(\sel\) and the move runs on the branches.
We detail the construction in the \emph{one-hot} realisation, whose
single-qubit controls make the mechanism transparent and which is
depth-tight at \(T\) selector qubits; the
pseudocode is in \cref{alg:stlcu}. The
\emph{dense} realisation (\cref{app:compress}) is memory-tight, with
selector \(\sum_k\lceil\log_2(m_k{+}1)\rceil\), but its multiplexed oracle
controls cost a \(\log\log(m{+}1)\) factor on the makespan
(\cref{prop:compress}).

\paragraph{Setup.}
A \emph{binning} is a fixed assignment of the \(T\) terms into \(g\)
blocks (bins) \(G_0,\dots,G_{g-1}\) of sizes \(m_0,\dots,m_{g-1}\)
(\(\sum_k m_k=T\)), fixed before the circuit is built; we argue over binnings
classically below. Block \(k\) holds terms
\((k,l)\), \(l=0,\dots,m_k{-}1\). Write
\(\alpha_k=\sum_l\lvert\alpha_{k,l}\rvert\),
\(\beta_k=\sqrt{\alpha_k/\alpha}\) (the block weight),
\(\gamma_{k,l}=\sqrt{\lvert\alpha_{k,l}\rvert/\alpha_k}\) (the term weight),
so \(\beta_k\gamma_{k,l}=\sqrt{\lvert\alpha_{k,l}\rvert/\alpha}\). The phase
\(e^{i\theta_{k,l}}\) rides the phased oracle \(e^{i\theta_{k,l}}U_{k,l}\).

Registers: \(g\) branch registers \(\branch_0,\dots,\branch_{g-1}\) of \(n\)
qubits each, all initially \(\ket{0}\) save \(\branch_0\) which holds the input
\(\ket{\psi}\); and a \emph{selector} that is a list of per-block one-hot
sub-registers \(\sel_k=(\sel_k[0],\dots,\sel_k[m_k{-}1])\), where
\(\sel_k[0]\) doubles as the block flag \(s_k\); the outer selector \(S\) is
\emph{absorbed} into the selector register. Activity is ``block \(k\) is
one-hot nonzero''; there is no reserved-zero, no separate \(S\). Total selector
width is \(T=\sum_k m_k\).

\begin{figure}[t]
\centering
\begin{tikzpicture}[
  >=Stealth, x=8mm, y=8mm,
  rwire/.style={line width=1.0pt, draw=black!55},
  dataflow/.style={line width=1.1pt, draw=accent},
  selflow/.style={line width=1.1pt, draw=select, line cap=round},
  selectline/.style={line width=0.9pt, draw=select, line cap=round},
  movearrow/.style={line width=1.2pt, draw=accent, >=Stealth},
  box/.style={draw, fill=boxfill, minimum height=5mm, minimum width=12mm,
             align=center, font=\small, inner sep=1pt},
  io/.style={font=\small},
  stagelbl/.style={font=\scriptsize, text=black!55},
]
\definecolor{accent}{HTML}{2E5C8A}
\definecolor{datacol}{HTML}{2E5C8A}
\definecolor{select}{HTML}{7A5C9E}
\definecolor{boxfill}{HTML}{E8E8E8}

\def\dyb{0.8}   
\def\dyB{1.6}   
\def\yFz{5.0}   
\pgfmathsetmacro{\yDz}{\yFz-\dyb}      
\pgfmathsetmacro{\yFo}{\yDz-\dyB}      
\pgfmathsetmacro{\yDo}{\yFo-\dyb}      
\pgfmathsetmacro{\yFt}{\yDo-\dyB}      
\pgfmathsetmacro{\yDt}{\yFt-\dyb}      

\def\xIn{0.0}
\def\xSpone{2.0}   
\def\xSptwo{3.2}   
\def\xButR{3.9}   
\def\xMoL{6.3}    
\pgfmathsetmacro{\xMoR}{\xMoL+1.5}    
\pgfmathsetmacro{\xMoC}{0.5*(\xMoL+\xMoR)}
\def\xOut{8.9}
\pgfmathsetmacro{\xMoveL}{\xButR+0.3}  
\pgfmathsetmacro{\xMoveM}{\xButR+0.9}  
\pgfmathsetmacro{\xMoveR}{\xButR+1.5}  
\pgfmathsetmacro{\yJ}{0.5*(\yFo+\yFt)}  

\draw[selectline] (\xIn,\yFz) -- (\xMoR,\yFz);
\draw[selectline] (\xButR,\yFo) -- (\xMoR,\yFo);
\draw[selectline] (\xButR,\yFt) -- (\xMoR,\yFt);
\draw[rwire] (\xIn,\yDz) -- (\xOut,\yDz);
\draw[rwire] (\xButR,\yDo) -- (\xOut,\yDo);
\draw[rwire] (\xButR,\yDt) -- (\xOut,\yDt);

\node[io, text=datacol] at ({\xIn-0.4},\yDz) {\(\branch_0\)};
\node[io, text=datacol] at ({\xIn-0.4},\yDo) {\(\branch_1\)};
\node[io, text=datacol] at ({\xIn-0.4},\yDt) {\(\branch_2\)};
\node[io, text=select] at ({\xIn-0.4},\yFz) {\(\sel_0\)};
\node[io, text=select] at ({\xIn-0.4},\yFo) {\(\sel_1\)};
\node[io, text=select] at ({\xIn-0.4},\yFt) {\(\sel_2\)};


\draw[selflow] (\xIn,\yFz) -- (\xButR,\yFz);                       
\draw[selflow] (\xSpone,\yFz) -- (\xSpone,\yJ) -- (\xSptwo,\yJ) -- (\xButR,\yFo);
\draw[selflow] (\xSptwo,\yJ) -- (\xButR,\yFt);
\fill[select] (\xSpone,\yFz) circle (1.8pt);
\fill[select] (\xSptwo,\yJ) circle (1.8pt);

\newcommand{\Xmark}[2]{
  \draw[movearrow] ({#1-\xh},{#2-\xh}) -- ({#1+\xh},{#2+\xh});
  \draw[movearrow] ({#1-\xh},{#2+\xh}) -- ({#1+\xh},{#2-\xh});%
}
\def\xh{0.16}  
\Xmark{\xMoveL}{\yDz}
\draw[movearrow] (\xMoveL,\yDz) -- (\xMoveL,\yJ) -- (\xMoveM,\yJ);
\fill[accent] (\xMoveM,\yJ) circle (1.6pt);
\draw[movearrow] (\xMoveM,\yJ) -- (\xMoveR,\yFo) -- (\xMoveR,\yDo);
\draw[movearrow, ->] (\xMoveR,{\yFo-0.5*\xh}) -- (\xMoveR,\yDo);
\Xmark{\xMoveR}{\yFo}
\draw[movearrow] (\xMoveM,\yJ) -- (\xMoveR,\yFt) -- (\xMoveR,\yDt);
\draw[movearrow, ->] (\xMoveR,{\yFt-0.5*\xh}) -- (\xMoveR,\yDt);
\Xmark{\xMoveR}{\yFt}

\def\bsp{0.13}    
\draw[selectline] (\xMoR,{\yFz-\bsp}) -- (\xOut,{\yFz-\bsp});
\draw[selectline] (\xMoR,\yFz)         -- (\xOut,\yFz);
\draw[selectline] (\xMoR,{\yFz+\bsp}) -- (\xOut,{\yFz+\bsp});
\draw[selectline] (\xMoR,{\yFo-\bsp}) -- (\xOut,{\yFo-\bsp});
\draw[selectline] (\xMoR,\yFo)         -- (\xOut,\yFo);
\draw[selectline] (\xMoR,{\yFo+\bsp}) -- (\xOut,{\yFo+\bsp});
\draw[selectline] (\xMoR,{\yFt-\bsp}) -- (\xOut,{\yFt-\bsp});
\draw[selectline] (\xMoR,\yFt)         -- (\xOut,\yFt);
\draw[selectline] (\xMoR,{\yFt+\bsp}) -- (\xOut,{\yFt+\bsp});

\node[box] at (\xMoC,\yFz) {Mött};
\node[box] at (\xMoC,\yFo) {Mött};
\node[box] at (\xMoC,\yFt) {Mött};

\node[io, text=select] at ({\xOut+0.4},\yFz) {\(\sel_0\)};
\node[io, text=select] at ({\xOut+0.4},\yFo) {\(\sel_1\)};
\node[io, text=select] at ({\xOut+0.4},\yFt) {\(\sel_2\)};


\node[stagelbl] at ({0.5*(\xIn+\xButR)},{\yDt-0.95}) {selector prep};
\node[stagelbl] at ({0.5*(\xMoveL+\xMoveR)},{\yDt-0.95}) {branch prep};
\node[stagelbl] at (\xMoC,{\yDt-0.95}) {Möttönen};

\end{tikzpicture}
\caption{The prep phase (\cref{sec:iterate}), \(g{=}3\). \emph{Selector prep}:
one-hot butterfly grows the branch flags \(\sel_k\) from a seed on
\(\sel_0\). \emph{Branch prep}: a mirroring CCNOT move tree routes the input
from \(\branch_0\) to the active branch. A per-branch Möttönen box builds the
term flags. Flag lines \(\sel_k\) above, data lines \(\branch_k\) below.}
\label{fig:prep}
\end{figure}
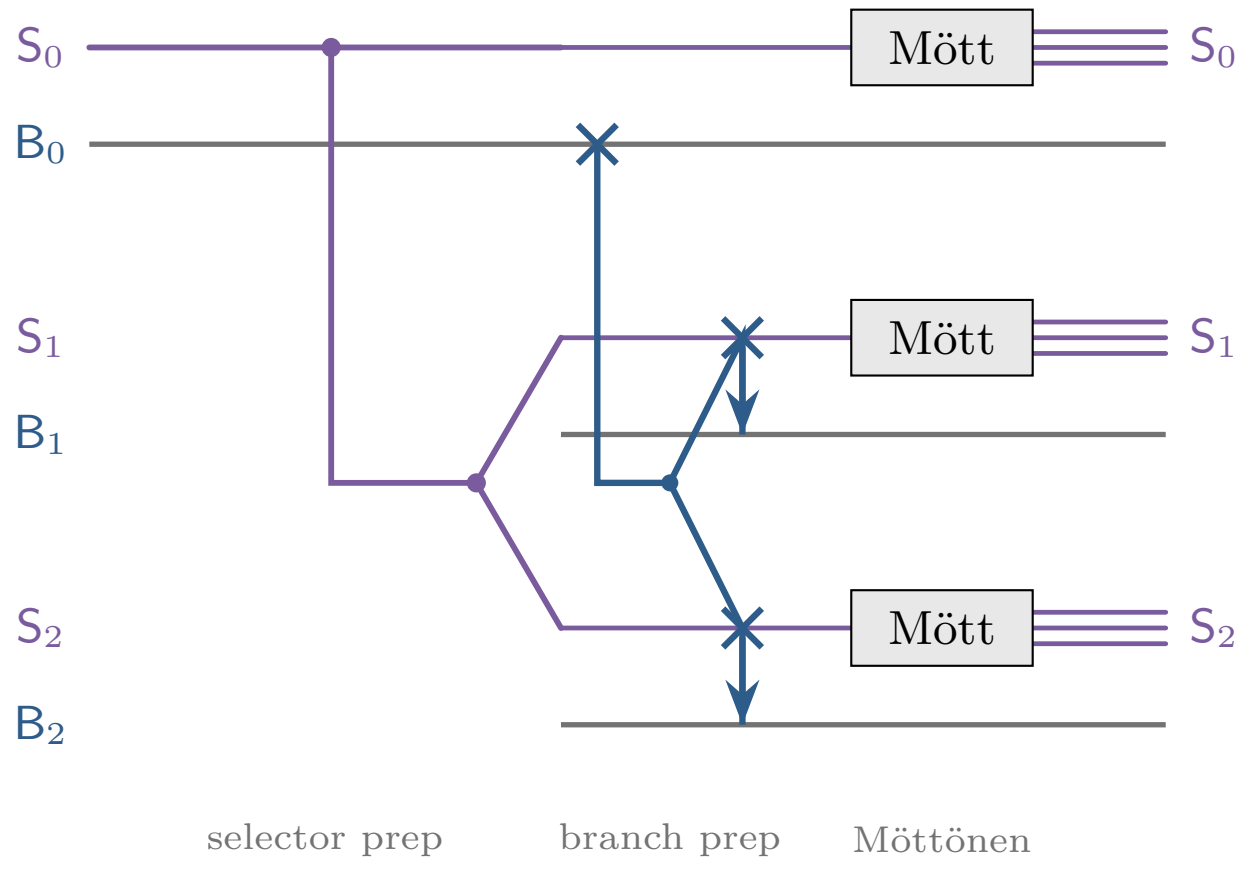

\paragraph{Prep, \cref{fig:prep}.}
The prep is \cref{alg:stlcu}, steps 0--5. An outer one-hot butterfly of
controlled-\(R_Y\)+CNOT gates prepares the block state
\(\sum_k\beta_k\ket{e_k}\) on the block flags \(\sel_k[0]\) in
\(\bigO{\log g}\). A one-qubit-per-block copy \(\sel\to\sel^{\mathrm{copy}}\)
then decouples the two parallel stages that follow: they act on disjoint
qubit sets (branch registers vs. selector flags) and share only the
block-active signal, so only the longer survives (a max, not a sum). On the
branches, a tree of controlled moves routes the input from \(\branch_0\) to
the active \(\branch_k\), fanning the path-decision flags \(W0\) (derived
off \(\sel^{\mathrm{copy}}\) by a level-parallel CX-merge) to the \(n\) data
qubits, costing \(\bigO{\log g+\log n}\). On the selector, an inner
self-gating butterfly splits each \(\sel_k[0]\) in place into the \(m_k\)
one-hot term flags, \(\bigO{\log m}\) (inactive blocks are a no-op). After
uncomputing \(W\) the state is
\(\sum_{k,l}\beta_k\gamma_{k,l}\ket{e_l}_{\sel_k}\ket{\psi}_{\branch_k}\ket{0}_{\text{rest}}\).
\Cref{app:thm-hybrid} gives the per-step mechanism.

\begin{figure}[t]
\centering
\begin{minipage}{0.97\columnwidth}
\small\raggedright
\noindent\textbf{Space--time LCU select}\quad
\(\mathrm{select}=\mathrm{prep}^{\dagger}\circ\mathrm{oracle}\circ\mathrm{prep}\)\\[4pt]
\textbf{Prep.}\\[1pt]
\(0.\) outer butterfly: prepare block state
\(\sum_k\beta_k\ket{e_k}\) on the block flags \(\sel_k[0]\)\hfill\((\bigO{\log g})\)\\
\(1.\) copy block flags \(\sel\to\sel^{\mathrm{copy}}\)
(decouples the two parallel stages)\\
\(2.\) \emph{move} \(\Vert\) \emph{inner}, on disjoint qubits; only the longer survives:\\
\quad move: tree-move input \(\branch_0\to\branch_k\), fan path flags
\(W0\to n\) data qubits\hfill\((\bigO{\log g+\log n})\)\\
\quad inner: split \(\sel_k[0]\) in place into the \(m_k\) one-hot term flags\hfill\((\bigO{\log m})\)\\
\(3.\) uncompute \(W0\); state is
\(\sum_{k,l}\beta_k\gamma_{k,l}\ket{e_l}_{\sel_k}\ket{\psi}_{\branch_k}\ket{0}_{\text{rest}}\)\\[3pt]
\textbf{Oracle.}\\[1pt]
\(4.\) apply phased \(e^{i\theta_{k,l}}U_{k,l}\) to \(\branch_k\) controlled on
\(\sel_k[l]\); parallel across \(k\)\hfill\((C)\)\\[3pt]
\textbf{Unprep.} (\(=\mathrm{prep}^{\dagger}\))\\[1pt]
\(5.\) re-derive \(W0\) off \(\sel^{\mathrm{copy}}\); reverse moves and butterflies;
clear \(\sel^{\mathrm{copy}}\)\\[3pt]
\textbf{Postselect.}\\[1pt]
\(6.\) measure \(\sel\); on \(\ket{0}^{\otimes T}\) yield
\((1/\alpha)A\ket{\psi}\) on \(\branch_0\)
\end{minipage}
\caption{The select as numbered steps. The dense variant replaces the inner
stage (step 2) by a Möttönen preparation at \(\bigO{m}\) (absorbed into
\(C\ge m\)) on \(\sum_k\lceil\log_2(m_k{+}1)\rceil\) qubits, paying a
\(\log\log(m{+}1)\) depth factor. Depth \(C+\Theta(\log(g\,\max\{n,m\}))\)
(one-hot) or \(\bigO{C\,\log\log(m{+}1)+\log(g\,\max\{n,m\})}\) (dense)
(\cref{thm:hybrid,prop:compress}); per-step mechanism in \cref{app:thm-hybrid}.}
\label{alg:stlcu}
\end{figure}

\paragraph{Oracle.}
Apply \(U_{k,l}\) to \(\branch_k\) controlled on \(\sel_k[l]=1\) (a single
qubit), followed by a controlled phase \(e^{i\theta_{k,l}}\) on the same flag.
The \(\branch_k\) are disjoint, so the \(g\) blocks run in parallel; within a
block the terms are sequential. The stage depth is the bin-packing makespan
\(C=\max_k\textstyle\sum_l d_{k,l}\) (the realised binning makespan).

\paragraph{Unprep.}
\(\mathrm{unprep}=\mathrm{prep}^{\dagger}\): reverse the inner and
outer butterflies and the tree-move (re-deriving \(W0\) off
\(\sel^{\mathrm{copy}}\) to guide the unmove), then copy-clear
\(\sel^{\mathrm{copy}}\). All routing ancilla return to \(\ket{0}\)
\emph{unconditionally} by the move-out symmetry of the move-in; only the \(T\)
selector flags carry failure.

\begin{figure}[t]
\centering
\begin{tikzpicture}[
  >=Stealth, x=9mm, y=9mm,
  rwire/.style={line width=1.0pt, draw=black!55},
  flow/.style={line width=1.1pt, draw=accent},
  selectline/.style={line width=0.9pt, draw=select},
  selflow/.style={line width=1.1pt, draw=select},
  gate/.style={draw, fill=gatefill, minimum height=6mm, minimum width=9mm,
              align=center, font=\small, inner sep=1pt},
  io/.style={font=\small},
  ctrldot/.style={circle, fill=select, inner sep=0pt, minimum size=1.7mm},
  stagelbl/.style={font=\scriptsize, text=black!55},
  brack/.style={dashed, gray},
]
\definecolor{accent}{HTML}{2E5C8A}
\definecolor{gatefill}{HTML}{E8E8E8}
\definecolor{select}{HTML}{7A5C9E}

\def\dy{1.5}        
\def\yB{3.6}        
\def\xIn{0.0}       
\def\xFanL{1.2}     
\def\xOraL{3.5}     
\def\gw{1.0}        
\def\ggap{0.2}      
\pgfmathsetmacro{\yBo}{\yB-\dy}
\pgfmathsetmacro{\yBt}{\yB-2*\dy}
\pgfmathsetmacro{\xDot}{\xOraL+0.5*\gw}        
\pgfmathsetmacro{\xGtwo}{\xDot+(\gw+\ggap)}          
\pgfmathsetmacro{\xGthree}{\xDot+2*(\gw+\ggap)}       
\pgfmathsetmacro{\xOraR}{\xOraL+2*(\gw+\ggap)+\gw}    
\pgfmathsetmacro{\xMid}{0.5*(\xOraL+\xOraR)}            
\pgfmathsetmacro{\xFanR}{2*\xMid-\xFanL}               
\pgfmathsetmacro{\xOut}{2*\xMid-\xIn}                   

\draw[rwire] (\xIn,\yB) -- (\xFanL,\yB);
\node[io] at ({\xIn-0.35},\yB) {\(\ket{\psi}\)};

\draw[flow] (\xFanL,\yB) -- (\xOraL,\yB);
\draw[flow] (\xFanL,\yB) -- (\xOraL,\yBo);
\draw[flow] (\xFanL,\yB) -- (\xOraL,\yBt);

\draw[rwire] (\xOraL,\yB)  -- (\xOraR,\yB);
\draw[rwire] (\xOraL,\yBo) -- (\xOraR,\yBo);
\draw[rwire] (\xOraL,\yBt) -- (\xOraR,\yBt);

\def\ctrlUp{0.55}   
\def\gateTop{0.34} 
\pgfmathsetmacro{\ySel}{\yB+\ctrlUp} 
\draw[selflow] (\xIn,{\ySel}) -- (\xFanL,{\ySel});
\node[io, text=select] at ({\xIn-0.35},{\ySel}) {\(\sel\)};
\draw[selflow] (\xFanL,{\ySel}) -- (\xOraL,{\ySel});          
\draw[selflow] (\xFanL,{\ySel}) -- (\xOraL,{\yBo+\ctrlUp}); 
\draw[selflow] (\xFanL,{\ySel}) -- (\xOraL,{\yBt+\ctrlUp}); 
\draw[selectline] (\xOraL,{\ySel})               -- (\xOraR,{\ySel});               
\draw[selectline] (\xOraL,{\yBo+\ctrlUp}) -- (\xOraR,{\yBo+\ctrlUp}); 
\draw[selectline] (\xOraL,{\yBt+\ctrlUp}) -- (\xOraR,{\yBt+\ctrlUp}); 
\draw[selectline] (\xDot,{\yB+\gateTop})  -- (\xDot,{\yB+\ctrlUp});
\node[ctrldot] at (\xDot,{\yB+\ctrlUp})  {};
\draw[selectline] (\xGtwo,{\yB+\gateTop}) -- (\xGtwo,{\yB+\ctrlUp});
\node[ctrldot] at (\xGtwo,{\yB+\ctrlUp}) {};
\draw[selectline] (\xDot,{\yBo+\gateTop})  -- (\xDot,{\yBo+\ctrlUp});
\node[ctrldot] at (\xDot,{\yBo+\ctrlUp})  {};
\draw[selectline] (\xGtwo,{\yBo+\gateTop}) -- (\xGtwo,{\yBo+\ctrlUp});
\node[ctrldot] at (\xGtwo,{\yBo+\ctrlUp}) {};
\draw[selectline] (\xGthree,{\yBo+\gateTop}) -- (\xGthree,{\yBo+\ctrlUp});
\node[ctrldot] at (\xGthree,{\yBo+\ctrlUp}) {};
\draw[selectline] (\xDot,{\yBt+\gateTop}) -- (\xDot,{\yBt+\ctrlUp});
\node[ctrldot] at (\xDot,{\yBt+\ctrlUp}) {};
\node[io, text=select] at ({\xOraR-0.25},{\yB+\ctrlUp+0.18})  {\(\sel_0\)};
\node[io, text=select] at ({\xOraR-0.25},{\yBo+\ctrlUp+0.18}) {\(\sel_1\)};
\node[io, text=select] at ({\xOraR-0.25},{\yBt+\ctrlUp+0.18}) {\(\sel_2\)};
\draw[selflow] (\xOraR,{\ySel})               -- (\xFanR,{\ySel});          
\draw[selflow] (\xOraR,{\yBo+\ctrlUp}) -- (\xFanR,{\ySel});          
\draw[selflow] (\xOraR,{\yBt+\ctrlUp}) -- (\xFanR,{\ySel});          
\draw[selflow] (\xFanR,{\ySel}) -- (\xOut,{\ySel});
\node[io, text=select, anchor=west] at ({\xOut+0.1},{\ySel}) {\(\to\ket{0}\)};
\newcommand{\Gate}[4]{
  \node[gate] at ({\xOraL + (#2)*(\gw+\ggap) + 0.5*\gw},#1) {\(U_{#3,#4}\)};}
\Gate{\yB}{0}{0}{0}  \Gate{\yB}{1}{0}{1}
\Gate{\yBo}{0}{1}{0} \Gate{\yBo}{1}{1}{1} \Gate{\yBo}{2}{1}{2}
\Gate{\yBt}{0}{2}{0}

\draw[brack] (\xOraL,{\yBt-0.55}) -- (\xOraL,{\yB+0.95});
\node[stagelbl] at (\xOraL,{\yB+1.15}) {prepared};
\draw[brack] (\xOraR,{\yBt-0.55}) -- (\xOraR,{\yB+0.95});
\node[stagelbl] at (\xOraR,{\yB+1.15}) {makespan};

\draw[flow] (\xOraR,\yB)  -- (\xFanR,\yB);
\draw[flow] (\xOraR,\yBo) -- (\xFanR,\yB);
\draw[flow] (\xOraR,\yBt) -- (\xFanR,\yB);

\draw[rwire] (\xFanR,\yB) -- (\xOut,\yB);
\node[io, anchor=west] at ({\xOut+0.1},\yB) {\(\propto A\ket{\psi}\)};

\node[stagelbl] at ({0.5*(\xFanL+\xOraL)},{\yBt-0.85}) {prep (distribute)};
\node[stagelbl] at ({0.5*(\xOraL+\xOraR)},{\yBt-0.85}) {oracle (parallel)};
\node[stagelbl] at ({0.5*(\xOraR+\xFanR)},{\yBt-0.85}) {unprep (collect)};

\end{tikzpicture}
\caption{One-hot space--time LCU, logical structure (\cref{sec:iterate}),
\(g{=}3\) bins of sizes \((2,3,1)\), \(T{=}6\). Prep distributes the
input across the branch registers; the oracle applies the controlled
\(U_{k,l}\) in parallel, gated by the one-hot flags \(\sel_k[l]\); unprep
(\(=\mathrm{prep}^{\dagger}\)) collects back to \(\branch_0\). Selector flow
(purple) mirrors data flow (blue); dashed brackets bound the makespan window
(slack below). Endpoints \(g{=}1\) sequential, \(g{=}T\) fully parallel.}
\label{fig:hierarchy}
\end{figure}
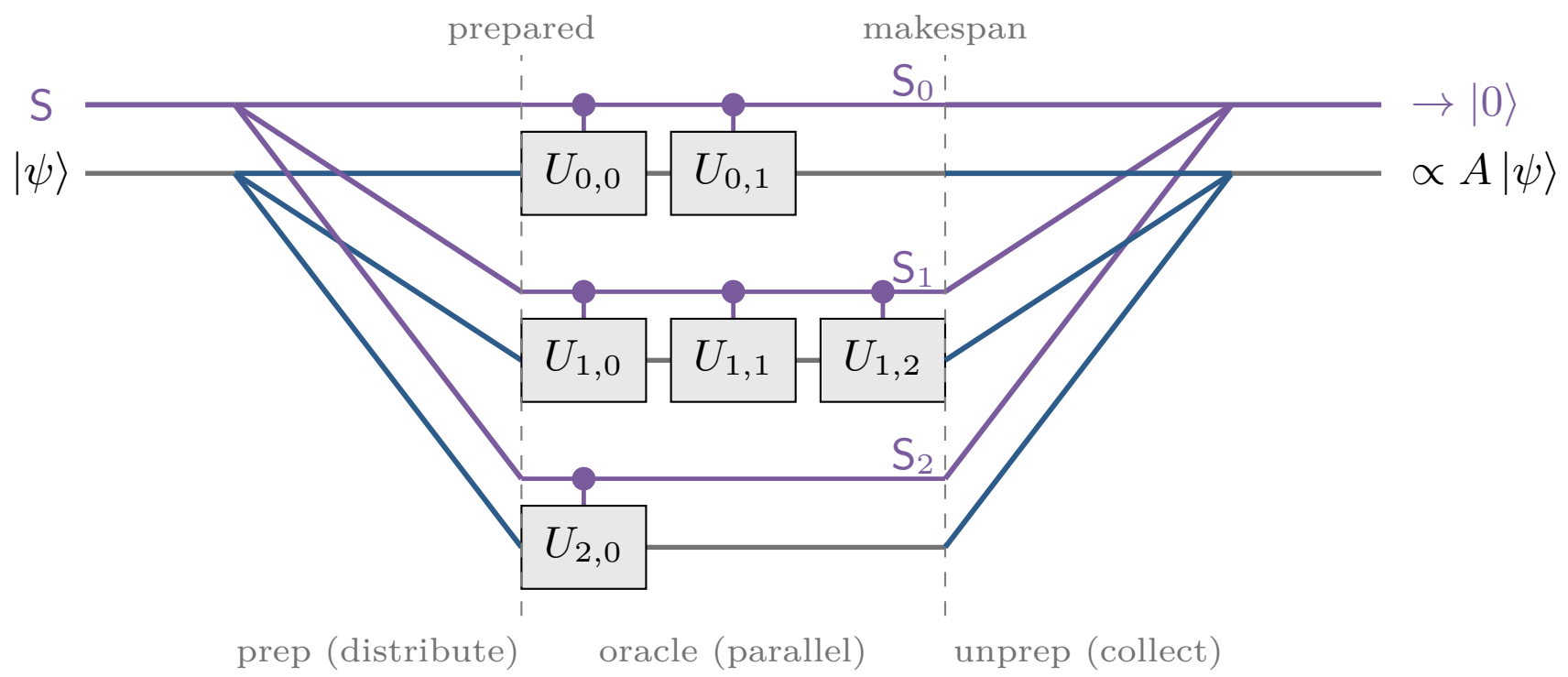

\begin{theorem}[Two-layer one-hot block-encoding]\label{thm:hybrid}
For \(T\) terms binned into \(g\) blocks of sizes \(m_0,\dots,m_{g-1}\)
(\(m=\max_k m_k\)), the construction of \cref{sec:iterate} is a
\(1/\alpha\)-block-encoding of \(A=\sum_j\alpha_j U_j\) with the standard LCU
success probability \(p_{\mathrm{succ}}=\alpha^{-2}\bra{\psi}A^\dagger A\ket{\psi}\):
on input \(\ket{\psi}\) the branch registers \(\branch_{\ne0}\) and all routing
ancilla return to \(\ket{0}\) unconditionally, and postselecting the selector
to \(\ket{0}^{\otimes T}\) produces
\((1/(\alpha\sqrt{p_{\mathrm{succ}}}))A\ket{\psi}\) on \(\branch_0\). Its depth and
width are
\[
  \begin{gathered}
  D=C+\bigO{\log(g\,\max\{n,m\})},\quad
  W(g)=\bigO{g\,n+T},
  \end{gathered}
\]
with \(C=\max_k\textstyle\sum_l d_{k,l}\) the binning makespan and selector
postselection on \(T\) qubits; peak width \((2g{-}1)n+T+g\) during the move
and \(gn+T+g\) during the oracle.
\end{theorem}

\begin{proof}[Proof sketch]
\emph{Correctness.} With \(\beta_{k,l}=\beta_k\gamma_{k,l}=\sqrt{\lvert\alpha_{k,l}\rvert/\alpha}\),
after prep the state is
\(\sum_{k,l}\beta_{k,l}\ket{e_l}_{\sel_k}\ket{\psi}_{\branch_k}\ket{0}_{\text{rest}}\);
after the oracle and the tree-unmove, the active branch's content is back on
\(\branch_0\) and every off-path register is \(\ket{0}\) (valid because the
one-hot selector keeps at most one branch live). The \(\sel=\ket{0}^{\otimes T}\)
component is
\(\sum_{k,l}\beta_{k,l}^2 e^{i\theta_{k,l}}U_{k,l}\ket{\psi}=(1/\alpha)A\ket{\psi}\)
(the amplitude \(\beta_{k,l}\) appears in prep and unprep and multiplies to
\(\lvert\alpha_{k,l}\rvert/\alpha\); the phase rides the controlled oracle), with
squared norm \(\alpha^{-2}\bra{\psi}A^\dagger A\ket{\psi}=p_{\mathrm{succ}}\).

\emph{Cost.} The data-routing chain (tree-move \(\bigO{\log g}\) plus per-bit
move fan \(\bigO{\log n}\), serialized) and the within-bin select (inner
butterfly \(\bigO{\log m}\)) act on disjoint qubit sets and share only the
block-active signal, so they run in parallel and only the longer survives;
the \(\sel\to\sel^{\mathrm{copy}}\) one-qubit-flag copy decouples them, holding
the block flags intact through the inner. Taking the max gives
\(\bigO{\log(g\,\max\{n,m\})}\); the oracle adds \(C\). Width peaks at
\((2g{-}1)n+T+g\) during the move (branches, transient routing fans, and the
\(g\)-qubit copy) and \(gn+T+g\) during the oracle.
\Cref{app:proofs} gives the full correctness trace and cost accounting.
\end{proof}

\paragraph{Fan-out model.}\label{par:fanout}
The depth bound holds in the bounded-fan-out model (\cref{sec:model}): each fan
is a balanced CNOT tree (\(\bigO{\log n}\)), the per-bit tree-move Toffoli reads
the fanned copies so no qubit controls two gates per layer, and the CX-merge is
level-parallel. Under unbounded fan-out every fan collapses to \(\bigO{1}\) and
the routing reduces to \(\bigO{\log g+\max\{\log g,\,\log m\}}\), independent of \(n\)
(the non-physical unbounded-fan-out idealisation of \cref{sec:bound}, in which
the per-bit move fan \(\log n\) disappears); the bounded-fan-out bound above
is the physical one.

\paragraph{Selector realisations.}
The construction above is in the \emph{one-hot} realisation: \(T\) flags, each
oracle controlled on a single qubit, depth-tight
\(C+\bigO{\log(g\,\max\{n,m\})}\) at the cost of \(T\)
postselected qubits. The
\emph{dense} (M\"ott\"onen) realisation replaces the \(T\) flags by per-block
reserved-zero registers \(R_k\) of \(\lceil\log_2(m_k{+}1)\rceil\) qubits
(block flag \(s_k\) folded in as seed), reaching the standard LCU footprint
\(\sum_k\lceil\log_2(m_k{+}1)\rceil\); its \(\bigO{m}\) prep is absorbed into
\(C\ge m\), but each oracle gate is controlled on \(\lceil\log_2(m{+}1)\rceil\)
qubits, so the makespan acquires a \(\log\log(m{+}1)\) factor, depth
\(C\log\log(m{+}1)+\bigO{\log(g\,\max\{n,m\})}\). It is memory-tight but not
depth-tight. Full statement and proof in
\cref{prop:compress,app:compress}; neither realisation subsumes the other
(one-hot pays memory for tight depth, dense pays a \(\log\log(m{+}1)\) depth factor for
tight memory), and the choice is independent of the scheduling regime (atomic
vs preemptive). The dense realisation is the standard LCU selector (dense
address register and multiplexed multi-controlled oracles), parallelised
over the \(g\) bins: at \(g{=}1\) its depth is \(\sum_j d_j\,\log\log T\), the
standard LCU select depth under bounded fan-out; its \(\log\log(m{+}1)\)
factor is the same bounded-fan-out light-cone cost as the routing term
(\cref{lem:routing,lem:withbin}), not a new one. The one-hot realisation is
strictly faster than standard LCU: single-qubit controls remove the
\(\log\log(m{+}1)\) factor, so its depth is never above the dense one and is
strictly below it for \(m\ge 4\) (the two coincide at \(m\le 3\), where the
multiplexed control is on at most two qubits).

\begin{table}[t]
\centering
\small
\begin{tabular}{lcc}
\toprule
 & depth & width \\
\midrule
floor & \(C^{*}{+}\Omega(\log(g\,\max\{n,m\}))\) & \(\Omega(gn{+}\textstyle\sum_k\lceil\log_2(m_k{+}1)\rceil)\) \\
one-hot & \(C{+}\Theta(\log(g\,\max\{n,m\}))\) \checkmark & \(\bigO{gn+T}\) \\
dense & \(\bigO{C\,\log\log(m{+}1)+\log(g\,\max\{n,m\})}\) & \(\Theta(gn{+}\textstyle\sum_k\lceil\log_2(m_k{+}1)\rceil)\) \checkmark \\
\bottomrule
\end{tabular}
\caption{The two selector realisations against the lower bound (\(\checkmark\)
marks the tight axis). Work term \(C\) is exact, matched by \(C^{*}\);
routing is \(\Theta\) where matched, \(\bigO{}\) otherwise. One-hot is
depth-tight at \(T\) selector qubits (\cref{thm:hybrid}); dense is
width-tight at the standard LCU footprint, paying an unmatched
\(\log\log(m{+}1)\) depth factor (\cref{prop:compress}). Neither reaches
both axes. Dense is standard LCU parallelised over \(g\) bins; one-hot is
never slower and strictly faster for \(m\ge 4\).}
\label{tab:tradeoff}
\end{table}

\begin{lemma}[Dense-selector bound]
\label{prop:compress}
The dense realisation of \cref{app:compress} (seeded fixed-point preparation)
is a \(1/\alpha\)-block-encoding of \(A\) with the standard LCU success
probability, for every \(g\in[1,T]\) and every binning, at
\[
  \begin{gathered}
  W(g)=\bigO{g\,n+\textstyle\sum_k\lceil\log_2(m_k{+}1)\rceil},\\
  D=\bigO{C\,\log\log(m{+}1)+\log(g\,\max\{n,m\})}.
  \end{gathered}
\]
with peak selector memory \(\sum_k\lceil\log_2(m_k{+}1)\rceil\) (the standard LCU
footprint, against one-hot's \(T\)) and peak width
\((2g{-}1)n+\sum_k\lceil\log_2(m_k{+}1)\rceil\) during the move;
\(C=\max_{k\in[g]}\sum_{j\in G_k}d_j\) and \(m=\max_k m_k\). It is
memory-tight but not depth-tight: its \(\bigO{m}\) Möttönen prep is absorbed
into \(C\ge m\), but each oracle gate is controlled on
\(\lceil\log_2(m{+}1)\rceil\) qubits, so the makespan acquires a
\(\log\log(m{+}1)\) factor (\cref{rem:ancilla}; \(1\) at \(m\le3\)).
\end{lemma}

\begin{proof}[Proof sketch]
\emph{Correctness.} Relabel the \(T\) term-addresses as pairs \((k,l)\) with
\(\beta_k\gamma_{k,l}=\sqrt{\lvert\alpha_{k,l}\rvert/\alpha}\); the dense selector
\(\sum_k\sum_l\beta_k\gamma_{k,l}\ket{R_k=l}\) is \cref{thm:hybrid}'s flat
selector in the reserved-zero factored basis, and the select is the
block-diagonal multiplex
\(\sum_{k,l}\ket{R_k=l}\!\bra{R_k=l}\otimes e^{i\theta_{k,l}}U_{k,l}\). Hence
\cref{thm:hybrid}'s trace applies verbatim, giving the \(R=\ket{0}\)
postselected component \((1/\alpha)A\ket{\psi}\) with the standard success
probability.

\emph{Cost.} The seeded fixed-point gadget (\cref{app:compress}) prepares
\(R_k\) in \(\bigO{m_k}\) per block, absorbed into \(C\ge m\). Each oracle gate
is controlled on \(c_k=\lceil\log_2(m_k{+}1)\rceil\) qubits; under bounded
fan-out a \(c\)-controlled gate costs \(\bigO{\log c}=\bigO{\log\log(m{+}1)}\)
depth (Barenco fan-in-2 tree, \cref{rem:ancilla}), so each oracle layer costs
\(d_{k,l}\,\bigO{\log\log(m_k{+}1)}\) and the makespan becomes
\(C\,\bigO{\log\log(m{+}1)}\) (clamped at \(1\) for \(m\le 3\)). The routing term is
the same disjoint-qubit max as one-hot. \Cref{app:proofs} gives the full
cost derivation and the borrowed-ancilla accounting.
\end{proof}

\begin{remark}[Borrowed ancilla for the multiplexed control]
\label{rem:ancilla}
The \(\bigO{\log(m{+}1)}\) control cost above is the ancilla-free
baseline (Barenco's \(O(c)\)-depth, \(0\)-ancilla decomposition of a
\(c\)-controlled gate). Borrowing \(O(c_k)\) ancilla per bin, a fan-in-2
tree of Toffoli gates at \(O(\log c_k)\) depth, drops the factor to
\(\bigO{\log\log(m{+}1)}\). This ancilla is of the same order as the
selector register \(R_k\) (\(O(c_k)=O(\log m_k)\) vs \(c_k=\lceil\log_2(m_k{+}1)\rceil\)),
so it is absorbed by the width bound
\(\bigO{gn+\textstyle\sum_k\lceil\log_2(m_k{+}1)\rceil}\) in big-\(O\); only the
constant factor of the footprint grows. We therefore take the dense depth as
\(C\,\log\log(m{+}1)+\bigO{\log(g\,\max\{n,m\})}\), width-tight in big-\(O\).
The in-place variant attains the information-theoretic footprint exactly, at
the cost of the larger \(\log(m{+}1)\) depth factor (ancilla-free), or
\(\log\log(m{+}1)\) with borrowed ancilla (\cref{rem:ancilla}).
\end{remark}

\paragraph{Heterogeneous depths: bin packing.}
Minimising \(D\) over the choice of bins is precisely the \(g\)-machine
\emph{makespan bin-packing} (multiprocessor scheduling) problem: assign jobs of
length \(d_j\) to \(g\) machines to minimise the maximum load.
It is NP-hard in general, but admits a polynomial-time
\((1+1/g)\)-approximation by longest-processing-time-first (LPT) scheduling and
a polynomial-time approximation scheme, so the assignment of terms to bins is a
well-studied combinatorial sub-problem rather than a new obstacle.

\section{Preemptive space--time LCU: cutting terms to reach the work floor}
\label{sec:schedule}

The atomic construction of \cref{thm:hybrid} treats each \(U_j\) as an
\emph{indivisible} \(n\)-wide black-box rectangle, so its depth is the
non-preemptive makespan \(C^*\)---the optimum of bin-packing. The work floor
\(\max\{\max_j d_j,\,(\sum_j d_j)/g\}\) of \cref{cor:preemptive-lb} sits below
\(C^*\) by the bin-packing slack. We close that gap by switching to the
\emph{preemptive} model, which assumes the stronger oracle model in which each
\(U_j\) is \emph{decomposable} into slices (a circuit oracle, not an opaque
black box). Cut each \(U_j\) into constant-weight slices, give each slice a
home register, and let slices migrate between registers in a preemptive
schedule (McNaughton's wrap-around rule). The atomic and preemptive regimes are
dual: atomic \(=\) bin-packing (non-preemptive makespan), preemptive \(=\)
scheduling (work floor). The composite weight
\(\beta_k\gamma_{k,l}=\sqrt{\lvert\alpha_{k,l}\rvert/\alpha}\) stays in the
selector and never moves, so each slice carries its term's weight unchanged.

The construction is a \emph{reduction} to the atomic one, not a new object:
treating each slice as a separate oracle reuses \cref{thm:hybrid} directly and
inherits its tightness. The only addition is a flag-controlled migration
between a slice's home register and its scheduled host---a controlled move
(source cleared) and its unmove, controlled on the source term's read-only flag
so the selector retains its weight and the postselection sees the layout it
expects. The migration is the same register-move primitive as the atomic
data-routing, so it costs \(\bigO{1}\) on the critical path (\cref{lem:migrate})
and the routing combines as the same max. The full construction---McNaughton's
rule, the one-hot selector layout, and the depth/width accounting---is in
\cref{app:preemptive}.

\begin{lemma}[Migration overhead]
\label{lem:migrate}
Under McNaughton's rule (each term cut at most once, \(g-1\) cuts), the
preemptive migration adds \(\bigO{1}\) depth on the critical path and only
transient \(\bigO{gn}\) width. \Cref{app:lem-migrate} gives the accounting,
including the dense-address case.
\end{lemma}

\begin{theorem}[Preemptive tightness]
\label{thm:preemptive}
The preemptive construction reaches the Brent floor
\(\max\{\max_j d_j,\,\textstyle\sum_j d_j/g\}\), so
\[
  D_{\mathrm{pre}}(g)
  \;=\; \max\Bigl\{\max_j d_j,\,\tfrac{1}{g}\textstyle\sum_j d_j\Bigr\}
  \;+\; \Theta\!\left(\log(g\,\max\{n,m\})\right),
\]
\(\Theta\)-tight to \cref{cor:preemptive-lb}: the lower bound is \cref{lem:induced}
(the Brent floor is the optimal \(g\)-machine preemptive makespan), and the
upper bound is the reduction of \cref{app:preemptive}---each slice an oracle,
\(g-1\) McNaughton cuts, migration overhead \(\bigO{1}\) on the critical path
(\cref{lem:migrate}), zero postselection overhead. McNaughton's \(g-1\) cuts
always suffice and cut each term at most once; restricting to \(\le g-2\)
preemptions makes optimisation NP-hard \parencite{Shchepin_2008}, so \(g-1\) is
the threshold of polynomial-time optimality.
\end{theorem}

This closes the tightness argument: the depth floor holds in the
\(k\)-local atomic-oracle model (robust to register reuse, measurement,
reset, and teleportation),
the non-preemptive slack is closeable constructively, and the preemptive
construction is \(\Theta\)-tight to the pointwise bound.

\section{Related work and positioning}

The standard LCU framework originates with \textcite{Childs_2012} and the
truncated-Taylor-series simulation of \textcite{Berry_2015}; the
\(1/\alpha\) block-encoding subnormalisation we inherit, in the standard
setting of \textcite{Gilyen_2019}, is that of these
works.

On the cost-reduction axis, several recent works exploit \emph{structure in the
\(U_j\)} --- commutativity, Pauli-ness, locality --- to beat the generic bound.
This construction assumes none and supplies the generic bound and its matching
lower bound. \textcite{Daskin_2026} lowers LCU \emph{width}
via a Birkhoff--von Neumann decomposition that cuts the number of terms and
hence the ancilla register. \textcite{DellaChiara_2025} (FOQCS-LCU) reduce select
depth for \emph{structured} (Pauli-string) Hamiltonians to a constant, via
Dicke-state preparation and parallel singly-controlled Pauli gates, at
\emph{linear} ancilla cost, exploiting Hamiltonian structure rather than width.
\textcite{Boyd_2023} parallelises LCU select by partitioning the terms into
\emph{commuting} groups and applying each group in parallel, with an
\(\bigO{\log n}\) qubit increase for a depth reduction. This is another
structure-exploiting instance of the same width-for-depth idea, complementary
to the generic construction here, which assumes nothing about commutation.
Boyd also uses adaptive circuits and teleportation for constant-depth Clifford,
the entanglement-assisted regime noted after \cref{thm:pointwise}.
\textcite{Zhang_2024} parallelise Hamiltonian simulation via parallel
quantum walks for uniform-structured (sparse, local, Pauli-sum) Hamiltonians,
reaching a doubly-logarithmic \(O(\log\log(1/\epsilon))\) depth
dependence on precision, structure-specific and complementary to the generic
select here. \textcite{Vasconcelos_2025} reduce ancilla \emph{overhead} of a single
block-encoding by approximately uncomputing all but one ancilla for reuse, a
space-time tradeoff in the opposite direction, freeing width rather than
trading it for select depth.
Our construction targets general (arbitrary-depth) \(U_j\) and trades width
for depth across the family, from \(\sum_j d_j\) at \(g{=}1\) (sequential) to
\(\max_j d_j+\bigO{\log(Tn)}\) at \(g{=}T\) (fully parallel). The structured
works are complementary: FOQCS and Boyd are tighter on their special cases,
and ours supplies the general width-for-depth tradeoff together with the
matching lower bounds of \cref{thm:pointwise,thm:routing}.
The binning \((g,\{G_k\})\) parameterizes this frontier explicitly
and identifies depth with the binning makespan; the operating point is
instance- and platform-dependent rather than universal. The closest in
spirit is the parallel quantum signal processing of
\textcite{Martyn_2025}, which parallelises by polynomial factorization;
that targets quantum signal processing rather than general LCU selects, and
the two are complementary.
On the lower-bound side, the work term is an instance of Brent's theorem
\parencite{Brent_1974} with \(g\) processors, and the routing term a
\(k\)-local light-cone cost in the spirit of the VLSI area--time bounds of
\textcite{Savage_1981}. The contribution here is that
both are tight for the quantum select in the \(k\)-local atomic-oracle model.

\section{Discussion: memory for depth}

Depth was the one block-encoding constraint with no knob: the select ran in
series, so depth was fixed at \(\sum_j d_j\). The \(g\)-branched family makes
it tunable, and the result is its tradeoff frontier, certified tight. Standard
ingredients (a controlled move, a one-hot butterfly, McNaughton's cutting rule)
make the select depth a scheduling makespan plus the matched logarithmic
routing, and the lower bound is recovered by lifting any realisation to the
schedule it induces: at width \(gn\) any realisation pays at least
\(C^*+\Omega(\log(g\,\max\{n,m\}))\) in the \(k\)-local atomic-oracle model, and
register reuse, mid-circuit measurement, reset, and teleportation do not reduce
it. The preemptive variant closes the bin-packing slack, so the work term
re-instantiates classical parallel complexity. The optimum is certified: no
routing wins depth for free, only the one log-factor the light cone charges.

What changes with \(g\) is the shape of the circuit, not its block-encoding
quality. The parallel construction pays \(g\) data registers to make the
controlled-unitary budget a maximum rather than a sum, while the \(1/\alpha\)
subnormalisation, success probability, and amplitude-amplification profile stay
those of standard LCU; the only non-unitary step is the postselection. For
heterogeneous-depth \(U_j\), locating the operating point reduces to a
well-studied scheduling problem with no universal optimum (\cref{fig:spacetime}).
The two selector realisations are tight on different axes: one-hot is
depth-tight (at \(T\) selector qubits), dense is memory-tight (the
information-theoretic footprint, at a \(\log\log(m{+}1)\) depth cost). That cost
is a double logarithm, so in practice the dense realisation is the natural
choice; the one-hot remains the exact witness that the floor is attainable.

\appendix
\section{Deferred proofs}
\label{app:proofs}

This appendix gives the full proofs sketched in \cref{sec:bound,sec:iterate}.

\subsection{Proof of \cref{lem:induced}}
\label{app:lem-induced}

The selector-controlled oracles act in superposition, but each \(U_j\)'s
gates still occupy definite layers of the circuit layout, so the bound below is
on layout width, not on branch amplitudes. For each call let
\(I_j\subset[0,D)\) be its layer interval; atomic calls are
contiguous of length \(d_j\), preemptive ones are unions of pieces totalling
\(d_j\). At most \(g\) intervals overlap any layer (the window is
\(n\)-wide, width is \(gn\), concurrent calls are disjoint). For the atomic
case the family has pointwise depth
\(\le g\), hence is \(g\)-colourable (interval graphs are optimally coloured
greedily with \(\omega\le g\) colours); a colour is a
machine, an interval coloured \(k\) is job \(j\) running non-preemptively on
machine \(k\), and its makespan \(\le D\), so \(D\ge C^*\). For the
preemptive case the same concurrency bound gives the work floor
\(\sum_j d_j/g\) (total work \(\sum_j d_j\), cleared at rate \(\le g\)), and the
causal ordering of one job's pieces (a job runs on at most one machine at a
time) gives \(\max_j d_j\); their max is the Brent floor, the classical
preemptive optimum (McNaughton's wrap-around rule \parencite{McNaughton_1959}
attains it; both terms are lower bounds in the \(k\)-local atomic-oracle model).
The width-floor clause \(\Omega(gn)\) is robust to register reuse, measurement,
reset, and teleportation: a running oracle's \(n\)-qubit intermediate state
occupies \(\ge n\) qubits for \(d_j\) layers wherever it is parked, and these
operations extend to a unitary on a larger (purified) system to which the same
window argument applies, so \(g\) overlapping windows still need \(\ge gn\)
qubits. The selector-floor clause \(\sum_k\lceil\log_2(m_k{+}1)\rceil\) is
robust for the same reason: the disjoint-gate constraint on the \(g\) concurrent
controls is a layout property that survives purification, and the per-branch
address space (\(m_k\) terms plus the reserved off-state) is unchanged.

\subsection{Proof of \cref{thm:pointwise}}
\label{app:thm-pointwise}

\Cref{lem:induced} gives \(D\ge C^*\). The prep precedes the makespan
branch's oracle (which reads the registers the prep writes), so it is
additive: \cref{lem:routing} gives \(D\ge C^*+\Omega(\log(gn))\) and
\cref{lem:withbin} gives \(D\ge C^*+\Omega(\log(gm))\). The two spans act on
disjoint registers (data vs. selector), so neither subsumes the other and they
combine by a max. The \(\Theta\)-equivalence is
\(\max\{\log(gm),\log(gn)\}=\log g+\max\{\log m,\log n\}=\log(g\,\max\{n,m\})\).
The construction (\cref{thm:hybrid}) realises any binning at
\(C+\bigO{\log(g\,\max\{n,m\})}\), so upper and lower bounds match up to
constants and the bin-packing slack.

\subsection{Proof of \cref{thm:bridge}}
\label{app:thm-bridge}

Upper bound \cref{thm:hybrid}:
\(D\le C(B)+\bigO{\log(g\,\max\{n,m\})}\). Lower bound \cref{thm:pointwise}:
\(D\ge C^*+\Omega(\log(g\,\max\{n,m\}))\). They match up to constants when
\(C(B)=C^*\). The lower bound is the reduction of \cref{lem:induced} (the work
floor) plus the routing of \cref{lem:routing,lem:withbin}: any width-\(gn\)
realisation induces a \(g\)-machine non-preemptive schedule of makespan
\(\le D\), hence \(D\ge C^*\), and the makespan branch's register conflict
adds the data-routing term.

\subsection{Proof of \cref{thm:hybrid}}
\label{app:thm-hybrid}

\emph{Correctness.} Let \(\beta_{k,l}=\beta_k\gamma_{k,l}=\sqrt{\lvert\alpha_{k,l}\rvert/\alpha}\).
After \(\mathrm{prep}\) the state is
\(\sum_{k,l}\beta_{k,l}\ket{e_l}_{\sel_k}\ket{\psi}_{\branch_k}\ket{0}_{\text{rest}}\)
(one selector flag set; input on the active branch). After the oracle,
\(\sum_{k,l}\beta_{k,l}\ket{e_l}_{\sel_k}(e^{i\theta_{k,l}}U_{k,l}\ket{\psi})_{\branch_k}\ket{0}_{\text{rest}}\);
after \(\mathrm{unprep}\)'s tree-unmove the active branch's content is back on
\(\branch_0\) and every off-path register is \(\ket{0}\) (valid precisely because
the one-hot selector keeps at most one branch live). The selector is intact
through the oracle (read as a control only), so \(\mathrm{unprep}\)'s inverse
butterflies re-merge it toward \(\ket{0}^{\otimes T}\): the
\(\sel=\ket{0}^{\otimes T}\) component is
\(\sum_{k,l}\beta_{k,l}^2 e^{i\theta_{k,l}}U_{k,l}\ket{\psi}=(1/\alpha)A\ket{\psi}\)
(the amplitude \(\beta_{k,l}\) appears twice, prep in and unprep out, and
multiplies to \(\beta_{k,l}^2=\lvert\alpha_{k,l}\rvert/\alpha\); the phase
\(e^{i\theta_{k,l}}\) rides the controlled oracle), with branch and routing
registers factored as \(\ket{0}\). Its squared norm is
\(\alpha^{-2}\bra{\psi}A^\dagger A\ket{\psi}=p_{\mathrm{succ}}\).

\emph{Cost.} The routing realises the two cost groups the lower bound
(\cref{sec:bound}) certifies as unavoidable, on disjoint qubit sets, so the
larger dominates (a \emph{max}, not a sum). Concretely: the data-routing
chain (the tree-move that routes the input to the active branch,
\(\bigO{\log g}\), and the per-bit move fan that fans the branch-selection
signal to the \(n\) data qubits it gates, \(\bigO{\log n}\)) is serialized
--- the move needs the fanned branch signal --- giving \(\bigO{\log g+\log n}\). The
within-bin select (the inner term selector, an in-place butterfly on
the selector flags, \(\bigO{\log m}\)) acts on disjoint qubits from the
data-routing chain and shares only the block-active signal, so it runs in
\emph{parallel} with it. A one-qubit-flag copy \(\sel\to\sel^{\mathrm{copy}}\)
holds the block flags intact through the inner (which destroys \(s_k[0]\)),
letting the path-decision flags be derived from \(\sel^{\mathrm{copy}}\) while
the inner runs on \(\sel\) and the move runs on the branches; all are
uncomputed before the oracle. The outer selector prep (butterfly and CX-merge
that derive those path-decision flags) adds a leading \(\bigO{\log g}\)
preceding the chain. The data-routing side is sequential in several \(\bigO{\log g}\)
phases --- outer butterfly, derive \(W0\), tree-move descent, uncompute
\(W0\) --- so its depth is \(\bigO{\log g}\) (the leading term absorbs them,
since each is \(\bigO{\log g}\)). Taking the max of the two groups gives
\(\bigO{\log g+\max\{\log g+\log n,\,\log m\}}\),
which simplifies to \(\bigO{\log(g\,\max\{n,m\})}\): the inner
\(\log g\) is dominated by the leading term, and the max contributes the
larger of \(\log n\) and \(\log m\).
The oracle contributes \(C\). Width: \(gn\)
branch registers, the \(T\)-qubit selector (postselected), the \(g\)-qubit
\(\sel^{\mathrm{copy}}\) (held live through the oracle, cleared in unprep),
and transient routing
(CX-merge storage \(\le g{-}1\), fanned \(W0\) \((g{-}1)(n{-}1)\), both cleared
before the oracle), peaking at \((2g{-}1)n+T+g\).

\subsection{Proof of \cref{prop:compress}}
\label{app:prop-compress}

\emph{Correctness.} Relabel the \(T\) term-addresses of \cref{thm:hybrid} as
pairs \((k,l)\), \(k\in[g]\), \(l\in[m_k]\), with
\(\beta_k=\sqrt{\alpha_k/\alpha}\),
\(\gamma_{k,l}=\sqrt{\lvert\alpha_{k,l}\rvert/\alpha_k}\),
\(\alpha_k=\sum_l\lvert\alpha_{k,l}\rvert\). The dense selector
\(\sum_k\sum_l\beta_k\gamma_{k,l}\ket{R_k=l,R_{\ne k}=0}\)
(exactly one \(R_k\) nonzero) is precisely \cref{thm:hybrid}'s flat selector in the
reserved-zero factored basis, since
\(\beta_k\gamma_{k,l}=\sqrt{\lvert\alpha_{k,l}\rvert/\alpha}\); the select is the
block-diagonal multiplex
\(\sum_{k,l}\ket{R_k=l}\!\bra{R_k=l}\otimes e^{i\theta_{k,l}}U_{k,l}\). Hence the
trace of \cref{thm:hybrid} applies verbatim
(\(\beta_j\rightsquigarrow\beta_k\gamma_{k,l}\)): the \(R=\ket{0}\) postselected
component is \((1/\alpha)A\ket{\psi}\) with the standard success probability.
(Equivalently: each bin is a sequential \(1/\alpha_k\)-block-encoding of
\(B_k=\sum_l\alpha_{k,l}U_{k,l}\); the outer gadget is a
\(1/\alpha\)-block-encoding of \(\sum_k B_k=A\) by \cref{thm:hybrid}; amplitudes
telescope, phases ride the phased oracles.)

\emph{Cost.} The dense realisation prepares \(R_k\) via the seeded fixed-point
gadget (\cref{app:compress}, \(\bigO{m_k}\) per block, parallel hence
\(\bigO{m}\)); this \(\bigO{m}\) prep is absorbed into the makespan
\(C\ge m\) (identity terms fold onto branch 0, so every binned term has depth
\(\Omega(1)\)). The select addresses \(U_{k,l}\) as a gate controlled on
\(c_k=\lceil\log_2(m_k{+}1)\rceil\) qubits (the multiplexed control). The address
\(l\) is in superposition, so the \(m\) terms' controls cannot share one
extracted flag: each term's control touches the data register, and a single
extract-to-flag would address only one term, not the multiplexed sum. Hence every
oracle layer pays the decomposed multi-control, not a one-off setup. Under
bounded fan-out, a \(c\)-controlled gate decomposes into \(\bigO{\log c}\)
Toffoli depth with \(\bigO{c}\) borrowed ancilla (Barenco fan-in-2 tree,
\cref{rem:ancilla}); this ancilla is of the same order as \(R_k\), so it is
absorbed by the width bound in big-\(O\). Thus each oracle layer
of depth \(d_{k,l}\) costs \(d_{k,l}\,\bigO{\log c_k}=d_{k,l}\,\bigO{\log\log(m_k{+}1)}\),
and the makespan becomes \(C\,\bigO{\log\log(m{+}1)}\) (\(\ge 1\), clamped at
single-qubit control for \(c_k\le 2\)). The preemptive migrate, by contrast,
runs one piece at a time, so its address is classical and a single extracted
flag suffices --- there the extract is a one-off (\cref{lem:migrate}). The routing
term is the same disjoint-qubit max as one-hot,
\(\bigO{\log(g\,\max\{n,m\})}\). Hence the dense depth is
\(D=C\,\log\log(m{+}1)+\bigO{\log(g\,\max\{n,m\})}\) at peak selector memory
\(\sum_k\lceil\log_2(m_k+1)\rceil\). Width is one \(n\)-qubit branch per bin
plus \(\bigO{gn}\) routing ancilla:
\(W(g)=\bigO{gn+\textstyle\sum_k\lceil\log_2(m_k{+}1)\rceil}\) (the selector is
subleading to \(gn\) when \(m\ll n\)).

\subsection{Proof of \cref{lem:migrate}}
\label{app:lem-migrate}

Each migrate's flag is fanned to \(n-1\) copies once in prep (a
\(\bigO{\log n}\) setup, paid once and swallowed by the routing term); with
the flag pre-fanned, the \(n\) per-bit moves are two Toffolis each, run in
parallel, so a migration costs \(\bigO{1}\). McNaughton's wrap lays the jobs
on one contiguous timeline and slices it into \(g\) chunks, so each cut
boundary is between two adjacent branches; a cut term's head sits at the start
of one branch and its tail at the end of the previous, so the migration graph
(branch \(\to\) branch, one edge per cut) is a subgraph of the branch path and
has degree \(\le 2\). A degree-\(\le2\) graph is a disjoint union of paths,
hence two-colourable, so all migrate-ins batch into \(\le2\) layers and all
migrate-backs into \(\le2\); the total migration depth on the critical path is
\(\bigO{1}\), an additive offset that does not multiply \(D^{*}\) and is
swallowed by the routing term. At most \(g-1\) terms migrate; the fan workspace
\((g{-}1)(n{-}1)\) is transient and absorbed by \(\bigO{gn}\). In the dense
realisation the multi-qubit address \(R_k\) is first extracted to one flag
(\(\bigO{\log m_k}\) AND-reduction, reused across the term's pieces) then
fanned and migrated identically; no multi-controlled migrate is needed, and
the migrated piece is steered by one qubit rather than an address.

\section{The preemptive reduction}
\label{app:preemptive}

This appendix gives the preemptive construction of \cref{sec:schedule} in full:
the one-hot selector layout, the McNaughton schedule-to-construction
correspondence, and the depth/width accounting.

\paragraph{Selector.}
The selector is presented in the one-hot layout of \cref{thm:hybrid} (the
dense realisation of \cref{prop:compress} composes via extract-and-fan, in
\cref{app:compress}): a per-block one-hot flag register \(\sel_k\) of \(m_k\)
qubits, with \(\sel_k=0^{\otimes m_k}\) inactive and \(\sel_k[l]=1\) addressing
term \((k,l)\); the composite amplitude on flag \((k,l)\) is
\(\beta_k\gamma_{k,l}=\sqrt{\lvert\alpha_{k,l}\rvert/\alpha}\).
Because the selector is one-block-hot (exactly one \(\sel_k\) nonzero), the
active block is the one whose flag register is nonzero, and parallel addressing
of a parked sector is controlled on the single source flag
\(\sel_k[l]\) directly, with no auxiliary location pointer (the all-zero
target register on the foreign branch is itself the parkable state). The
flags are read-only controls throughout (the migrate copies/clears branch
data, not flags), so the selector retains weight \(\beta_k\gamma_{k,l}\) intact
and the content returns home.

\paragraph{Reduction.}
Take a preemptive schedule on \(g\) registers that cuts \(U_{k,l}\) into
\(p_{k,l}\) contiguous pieces and assigns each piece a home, packing pieces to
(near) the work floor. Each piece runs controlled on its term's address
\(R_k{=}l\); between pieces a controlled move migrates the \(n\)-qubit state to
the next home; after the last piece of \(U_{k,l}\) an \emph{unmove} returns the
content to \(U_{k,l}\)'s home register. The composite weight
\(\beta_k\gamma_{k,l}\) is carried in the selector \(R_k\) and never moves;
only the content migrates and returns, so the standard merge,
\(\mathrm{Prepare}^\dagger\), and postselection of \cref{sec:gadget} see the
layout they expect. All routing is unitary and selector/flag-controlled, hence
invisible to the postselection: the \(1/\alpha\) block-encoding and standard
success probability are preserved without degradation.

\begin{figure}[t]
\centering
\begin{tikzpicture}[
  >=Stealth, x=9mm, y=9mm,
  rwire/.style={line width=1.0pt, draw=black!55},
  ujfill/.style ={draw, fill=gray!18},
  ujBfill/.style={draw, fill=gray!40},
  regA/.style ={draw, fill=accent!22},
  regAmig/.style={draw, fill=accent!48},
  regB/.style ={draw, fill=select!22},
  regBmig/.style={draw, fill=select!48},
  io/.style={font=\small},
  ax/.style={font=\scriptsize, text=black!65},
  boxh/.style={font=\small},
]
\definecolor{accent}{HTML}{2E5C8A}
\definecolor{select}{HTML}{7A5C9E}
\newcommand{\DepthBox}[5]{%
  \draw[#1] (#2,#4-0.4) rectangle (#2+#3,#4+0.4);
  \node[boxh] at (#2+#3/2,#4) {#5};%
}

\def\yA{3.0}
\def\yB{0.8}
\def\Dstar{5.0}   
\def\ymid{1.9}    

\draw[rwire] (0,\yA) -- (\Dstar,\yA);
\draw[rwire] (0,\yB) -- (\Dstar,\yB);
\node[io] at (-0.3,\yA) {\(\branch_A\)};
\node[io] at (-0.3,\yB) {\(\branch_B\)};

\DepthBox{regAmig}{0}{1}{\yA}{$U_2^H$}
\DepthBox{regA}{1}{4}{\yA}{$U_1$}
\DepthBox{regB}{0}{3}{\yB}{$U_0$}
\DepthBox{regBmig}{3}{2}{\yB}{$U_2^T$}

\draw[->, thick, accent] (1,\yA-0.4) -- (1,\ymid)
  -- (3,\ymid) node[ax, midway, above] {migrate}
  -- (3,\yB+0.4);

\end{tikzpicture}
\caption{The preemptive move (\cref{sec:schedule}), schedule view. \(U_2\)
is cut at a McNaughton boundary into head \(U_2^H\) (\(\branch_A\)) and
tail \(U_2^T\) (\(\branch_B\)); the migrate arrow traces the job across the
cut (an unmove returns the state home). At most \(g-1\) cuts attain the work
floor; the migrate is controlled on the read-only flag \(\sel_k[l]\), so the
selector retains its weight \(\beta_k\gamma_{k,l}\).}
\label{fig:migrate}
\end{figure}
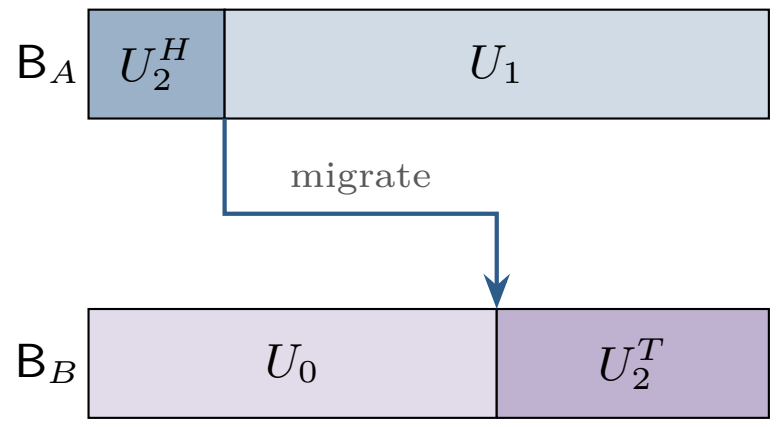

By McNaughton's wrap-around rule \parencite{McNaughton_1959}, laying the
jobs on a single timeline and slicing it into \(g\) contiguous chunks
attains the floor \(D^*=\max\{\max_j d_j,\,\sum_j d_j/g\}\); since
\(D^*\ge\max_j d_j\ge d_j\), every job fits in one chunk and straddles at most
one boundary, so each term is cut into at most two pieces (head and tail).
View each piece as a separate oracle; the split oracles are still unitaries,
so the atomic construction (\cref{thm:hybrid}) applies directly and inherits
its tightness. The only addition is the migration between a piece's home
register and its temporary host (\cref{fig:migrate}): a flag-controlled move
(source cleared) and its unmove, controlled on the source term's read-only flag
\(\sel_k[l]\) so the selector retains its weight \(\beta_k\gamma_{k,l}\) and the
postselection sees the layout it expects. The migration cost on the critical
path is \(\bigO{1}\) by \cref{lem:migrate}, and the routing is the same two
groups as the atomic case combined by the same max, the migration fan replacing
the per-bit move fan at \(\bigO{\log n}\). Thus
\[
  \begin{aligned}
  D_{\mathrm{sched}}(g)
  &\;=\; \max\Bigl\{\max_j d_j,\,\tfrac{1}{g}\textstyle\sum_j d_j\Bigr\}\\
  &\quad{}+\; \bigO{\log(g\,\max\{n,m\})},
  \end{aligned}
\]
\(\Theta\)-tight to \cref{cor:preemptive-lb} (for balanced binning
\(m\approx T/g\), this is the floor \(+\,\bigO{\log(Tn)}\)). Width is
\(W=\bigO{g\,n}\), the selector sub-leading for \(n\gg 1\). Cuts pay off
exactly in the uneven-binning regime where the atomic slack is nonzero; for
even binning no cuts are needed.

\section{The dense selector realisation}
\label{app:compress}

This appendix gives the \emph{dense} realisation, which is memory-tight:
the \(T\) term flags are replaced by a per-block reserved-zero
register \(R_k\) of \(\lceil\log_2(m_k+1)\rceil\) qubits (address
\(0\)=\emph{inactive}, address \(l{+}1\)=term \(l\)), with the branch flag
\(s_k\) consumed as \(R_k\)'s activation seed (address \(0\)=inactive is the
\emph{off} state, so the active-bin signal is \(R_k\ne0\) and no separate outer
one-hot survives). Its peak selector memory is
\(\sum_k\lceil\log_2(m_k+1)\rceil\), the standard LCU footprint, against the
one-hot variation's \(T\). The main construction (\cref{sec:iterate},
\cref{fig:hierarchy}) is presented there in the one-hot realisation, whose
single-qubit controls make the mechanism transparent.

\paragraph{Preparation (seeded fixed-point gadget).}
The block flag \(s_k\) is folded into \(R_k\) as the seed qubit (no separate
\(S\) register), which is what attains the peak
\(\sum_k\lceil\log_2(m_k+1)\rceil\) exactly. The term superposition is built
by a unitary \(U\) on \(R_k\) with the fixed-point action
\[
  U\ket{0\dots0}=\ket{0\dots0},\qquad
  U\ket{\mathrm{seed}{=}1}=-\sum_l\gamma_{k,l}\ket{l+1},
\]
so inactive blocks (seed \(0\)) are a no-op and the seed expands to the block
superposition over addresses \(1,\dots,m_k\). The global \(-1\) is a per-block
phase that cancels in the
\(\mathrm{prep}\cdot\mathrm{oracle}\cdot\mathrm{prep}^\dagger\) sandwich (it
appears in \(U\) and again in \(U^\dagger\)), so the realised operator is
exactly \(A/\alpha\).
Such a fixed-point preparation is a standard Möttönen state-preparation unitary
(a Householder reflection whose action is the one above); we use the
construction of \textcite{Mottonen_2005}, which prepares a nonnegative state on
\(c_k=\lceil\log_2(m_k{+}1)\rceil\) qubits in \(\bigO{m_k}\) depth with no
ancilla, sign-exact since the target is nonnegative. Across the \(g\) blocks
the preparation is \(\bigO{m}\), absorbed into the makespan \(C\ge m\). The
multiplexed oracle control is not absorbed: each gate is controlled on
\(\lceil\log_2(m{+}1)\rceil\) qubits, giving the \(\log\log(m{+}1)\) depth
factor of \cref{prop:compress}. The outer layer and input routing are the main
text's two-layer tree-move (\cref{sec:gadget}) throughout; the block-encoding,
success probability, and subnormalisation are identical to the one-hot
variation's (\cref{prop:compress}).

\printbibliography

\end{document}